\documentclass[twocolumn]{aastex63}
\pdfoutput=1
\usepackage{amsmath}	
\usepackage{xcolor}
\defcitealias{Riess2016}{R16}
\defcitealias{Hoffmann2016}{H16}

\newcommand{\hunit}{km s$^{-1}$Mpc$^{-1}$}

\shorttitle{Cepheid distance to the SNIa host galaxy NGC 5584}
\shortauthors{Javanmardi et al.}

\begin{document}

\title{Inspecting the Cepheid distance ladder:\\The \textit{Hubble Space Telescope} distance to the SNIa host galaxy NGC 5584}

\correspondingauthor{Behnam Javanmardi}
\email{behnam.javanmardi@obspm.fr, behjava@gmail.com}

\author[0000-0002-9317-6114]{Behnam Javanmardi}
\affiliation{LESIA, Observatoire de Paris, Université PSL, CNRS, Sorbonne Université, Université de Paris, 5 place Jules Janssen, 92195 Meudon, France}

\author{Antoine M\'erand}
\affiliation{European Southern Observatory, Karl-Schwarzschild-Str. 2, 85748 Garching, Germany}

\author[0000-0003-0626-1749]{Pierre Kervella}
\affiliation{LESIA, Observatoire de Paris, Université PSL, CNRS, Sorbonne Université, Université de Paris, 5 place Jules Janssen, 92195 Meudon, France}

\author[0000-0003-3889-7709]{Louise Breuval}
\affiliation{LESIA, Observatoire de Paris, Université PSL, CNRS, Sorbonne Université, Université de Paris, 5 place Jules Janssen, 92195 Meudon, France}

\author[0000-0001-7853-4094]{Alexandre Gallenne}
\affiliation{Nicolaus Copernicus Astronomical Centre, Polish Academy of Sciences,  Bartycka 18, 00-716 Warszawa, Poland}
\affiliation{Universidad de Concepci\'on, Departamento de Astronom\'ia, Casilla 160-C, Concepci\'on, Chile}
\affiliation{Unidad Mixta Internacional Franco-Chilena de Astronom\'ia (CNRS UMI 3386), Departamento de Astronom\'ia, Universidad de Chile,\\ Camino El Observatorio 1515, Las Condes, Santiago, Chile}

\author[0000-0002-7399-0231]{Nicolas Nardetto}
\affiliation{Universit\'e C\^ote d'Azur,  Observatoire de la C\^ote d'Azur, CNRS, Laboratoire Lagrange, France}

\author{Wolfgang Gieren}
\affiliation{Universidad de Concepci\'on, Departamento de Astronom\'ia, Casilla 160-C, Concepci\'on, Chile}

\author[0000-0002-9443-4138]{Grzegorz Pietrzy\'nski}
\affiliation{Nicolaus Copernicus Astronomical Centre, Polish Academy of Sciences,  Bartycka 18, 00-716 Warszawa, Poland}
\affiliation{Universidad de Concepci\'on, Departamento de Astronom\'ia, Casilla 160-C, Concepci\'on, Chile}

\author{Vincent Hocd\'e}
\affiliation{Universit\'e C\^ote d'Azur,  Observatoire de la C\^ote d'Azur, CNRS, Laboratoire Lagrange, France}

\author[0000-0002-4829-4955]{Simon Borgniet}
\affiliation{LESIA, Observatoire de Paris, Université PSL, CNRS, Sorbonne Université, Université de Paris, 5 place Jules Janssen, 92195 Meudon, France}




\begin{abstract}

The current tension between the direct and the early Universe measurements of the Hubble Constant, $H_0$, requires detailed scrutiny of all the data and methods used in the studies on both sides of the debate. The Cepheids in the type Ia supernova (SNIa) host galaxy NGC~5584 played a key role in the local measurement of $H_0$. The SH0ES project used the observations of this galaxy to derive a relation between Cepheids' periods and ratios of their amplitudes in different optical bands of the Hubble Space Telescope (HST), and used these relations to analyse the light curves of the Cepheids in around half of the current sample of local SNIa host galaxies. In this work, we present an independent detailed analysis of the Cepheids in NGC~5584. We employ different tools for our photometric analysis and a completely different method for our light curve analysis, and we do not find a systematic difference between our period and mean magnitude measurements compared to those reported by SH0ES. By adopting a period-luminosity relation calibrated by the Cepheids in the Milky Way, we measure a distance modulus $\mu=31.810\pm0.047$ (mag) which is in agreement with $\mu=31.786\pm0.046$ (mag) measured by SH0ES. In addition, the relations we find between periods and amplitude ratios of the Cepheids in NGC~5584 are significantly tighter than those of SH0ES and their potential impact on the direct $H_0$ measurement will be investigated in future studies.
\end{abstract}

\keywords{cosmology: distance scale -- methods: data analysis -- methods: observational -- galaxies: distances and redshifts -- stars: variables: Cepheids  -- techniques: photometric}



\section{Introduction}\label{sec:intro}
The current expansion rate of the Universe, known as the Hubble constant or $H_0$, is one of the fundamental parameters of the standard model of cosmology and of any viable cosmological model. A few decades after the initial estimate of around 500 km s$^{-1}$Mpc$^{-1}$ by \citet{Hubble1929}, $H_0$ became a place for debate with values either $\approx 100$ \hunit \citep[e.g. ][]{vandenBergh1970,deVaucouleurs1972} or $\approx 50$ \hunit \citep[e.g.][]{Sandage1975}. The debate was finally settled after the findings of the HST $H_0$ Key Project whose final results was $H_0 = 72 \pm 8$ \hunit \citep{Freedman2001}. This value was found to be in agreement with the subsequent results from the observations of the Cosmic Microwave Background (CMB) by the Wilkinson Microwave Anisotropy Probe \citep[WMAP, e.g. ][]{Spergel2003} based on the standard Lamda-Cold-Dark-Matter ($\Lambda$CDM) model. However, in the recent years and with the improved precision of the measurements of $H_0$, a significant tension has again risen this time between the so called early-Universe cosmology-dependent approaches finding $H_0 \approx 67$ \hunit and late-Universe direct measurements mostly finding $H_0 \approx 73$ \hunit.

On the one hand, using the precise observations of the CMB from the Planck satellite, \citet{Planck2018} concluded that the $\Lambda$CDM provides an excellent explanation of the CMB data and reported a model-dependent prediction of the Hubble constant $H_0=67.4 \pm 0.5$ \hunit. This result is in good agreement with other early Universe measurements. For example, by combining baryon acoustic oscillation (BAO) and Big Bang nucleosynthesis (BBN) data, \citet{Addison2018} reported a CMB-independent value of $H_0=66.98 \pm 1.18$ \hunit, and by combining BBN and BAO data with galaxy clustering and weak lensing data, the Dark Energy Survey reported $H_0=67.4^{+1.1}_{-1.2}$ \hunit \citep{Abbott2018DES}. In a recent study and based on new high resolution CMB observations from the Atacama Cosmology Telescope, \citet{Aiola2020ACT} reported $H_0=67.9 \pm 1.5$ \hunit, consistent with previous early Universe results. All these results are based on the $\Lambda$CDM model. 

On the other hand, the SH0ES (Supernovae $H_0$ for the Equation of State) project, that uses the Cepheid calibrated SNeIa data, finds a significantly higher locally measured $H_0$ value. Cepheids are one of the most reliable distance indicators and, using more than a decade of observations\citep[see e.g.][]{Riess2005}, the SH0ES project measures $H_0=73.5 \pm 1.4$ \hunit for the Hubble constant \citep{Riess2019ApJ, Riess2019Nat}. This is one of the most precise determinations of $H_0$ and is in more than $4\sigma$ tension with the early-Universe results. 

There are also other direct but Cepheid-independent methods for measuring $H_0$. Using time-delay data of gravitationally lensed quasars, the H0LiCOW (H0 lenses in COSMOGRAIL's Wellspring) project reported $H_0=73.3^{+1.7}_{-1.8}$ \hunit \citep{Wong2020}. In another gravitational lensing study, \citet{Birrer2020} employed a different lens mass profile modeling approach and found $H_0=67.4^{+4.1}_{-3.2}$ \hunit and $H_0=74.5^{+5.6}_{-6.1}$ \hunit using two different data sets. Using yet another independent method of geometric distance measurements to megamaser-hosting galaxies, \citet{Pesce2020} reported $H_0=73.9 \pm 3.0$ \hunit. Another late Universe method to measure the $H_0$ is to calibrate the SNeIa with the tip of the red giant branch (TRGB), using which the Carnegie-Chicago Hubble Program (CCHP) found $H_0=69.8 \pm 1.4$ \hunit \citep{Freedman2019}. 

Although some late Universe results are in agreement with those from early Universe approaches, the absolute majority of direct methods find $H_0$ values larger than the early Universe model-dependent methods, and currently different combinations of the late Universe measurements are in 4 to $6\sigma$ tension with the early Universe $\Lambda$CDM predictions \citep{Verde2019Nat, Riess2019Nat}, and removing any one method does not appear to resolve the tension. In fact, the general consistency of the direct methods on the one hand, and of those of the early-Universe methods on the other hand, significantly reduces the possibility that systematics in one method, data, or analysis would solve this problem.

However, since the persistence of the $H_0$ tension would mean the failure of the base $\Lambda$CDM model, and given the generally understood success of the $\Lambda$CDM in explaining the CMB and the large-scale structure data, it is absolutely necessary to not only take different approaches for measuring $H_0$, but also, as emphasized by \citet{Riess2020}, to scrutinize in detail all the data, methods, and the studies that have led to this cosmic discordance. 

The main three rungs of the Cepheid distance ladder are i) calibration of the period-luminosity relation using geometric distance measurements to nearby Cepheids, ii) calibration of the SNIa absolute magnitude using Cepheids in  SNIa host galaxies out to around 40 Mpc, and iii) using SNIa out to redshift 0.15 to measure $H_0$. 

On the first rung, different studies \citep[e.g.][]{Nardetto2018,Borgniet2019,Kervella2019b,Gallenne2019, Anderson2019, Hocde2020a,Hocde2020b,Musella2020} have focused on enhancing our understanding of the different properties of Cepheid variables, and in a recent study, \citet{Breuval2020} presented a new calibration of PL relation for Milky Way Cepheids using their companion parallaxes from Gaia \citep{Kervella2019}. In addition, the high precision measurement of the distance to the Large Magellanic Cloud (LMC) by \citet{Pietrzynski2019Nat} provides an accurate calibration of the Cepheid period-luminosity relation in this satellite galaxy of the Milky Way.

On the third rung, the impact of SNeIa environment \citep{Roman2018} and in particular that of the star formation rate \citep{Rigault2015} of their host galaxies on distance measurements has been investigated. \citet{Jones2018}, however, concludes that the environmental dependency of SNeIa properties have negligible effect on the $H_0$ measurements. \citet{Dhawan2018} and \citet{Burns2018} used near-infrared observations, where SNeIa luminosity variations and extinction by dust are less than in the optical observations, and concluded that the $H_0$ tension is likely not caused by systematics like dust extinction or SNeIa host galaxy mass. Also, \citet{Hamuy2020} reported that different methods for standardization of SNeIa light curves yield consistent results with a small standard deviation, concluding that SNeIa are robust calibrators of the third rung. 

In this study, we scrutinize the second rung, that is, the Cepheid calibration of SNeIa host galaxies. This intermediate rung plays the important role of connecting the geometric distance calibrations of the ladder to the cosmic scales. Therefore, it is vital to also investigate the observations and analysis involved in the second rung of the Cepheid distance ladder independently of SH0ES that has so far been the only program undertaking this effort. \citet[][hereafter R16]{Riess2016} used HST data to measure the Cepheid distances to 19 SNIa host galaxies for the calibration of the luminosity of larger distance SNIa. \citetalias{Riess2016} presents the near-infrared (NIR) observations, whereas a companion paper by \citet[][hereafter H16]{Hoffmann2016} reports the complete optical observations of Cepheid variables in these SNIa host galaxies. Out of these 19 galaxies, 10 have been observed in the earlier stages of the SH0ES project \citep{Riess2009} and with older HST instruments, i.e. NICMOS\footnote{Near Infrared Camera and Multi-Object Spectrometer.} and WFPC2\footnote{Wide Field and Planetary Camera 2.}, using F555W (V), F814W (I), and F160W (H) bands for measuring mean magnitudes and periods of their Cepheid variables\footnote{In this paper F555W, F814W, and F160W are used interchangeably with V, I, and H, respectively.}. However, for 9 out of 19 of these galaxies, the photometric time series necessary for identifying the Cepheids, measuring their light curves and estimating their periods, have been obtained using the wide band HST F350LP filter available on WFC3/UVIS. The wavelength range of this so called "white light" filter covers those of V and I bands, hence is suitable for the detection of faint sources. The 9 galaxies mentioned above currently have much fewer random phase observations in V and I bands, not sufficient for light curve analysis. NGC~5584, however, has time series observations in all these three bands. Using the data of this galaxy, the SH0ES team obtained a relation between the periods of the Cepheids and the ratio of their amplitudes in V and I relative to those in F350LP band. Then, assuming that these relations derived from NGC~5584 also hold in other SNIa host galaxies, the SH0ES project correct for the effect of random phase observations of Cepheids in the galaxies with few V and I observations. Therefore, the Cepheids in the galaxy NGC~5584 played a key role in obtaining the periods and mean magnitudes of the Cepheids in almost half of the current SH0ES sample of SNeIa host galaxies, and in turn in the final measurement of $H_0$. 

\begin{figure*}
    \centering
    \includegraphics[width=\textwidth]{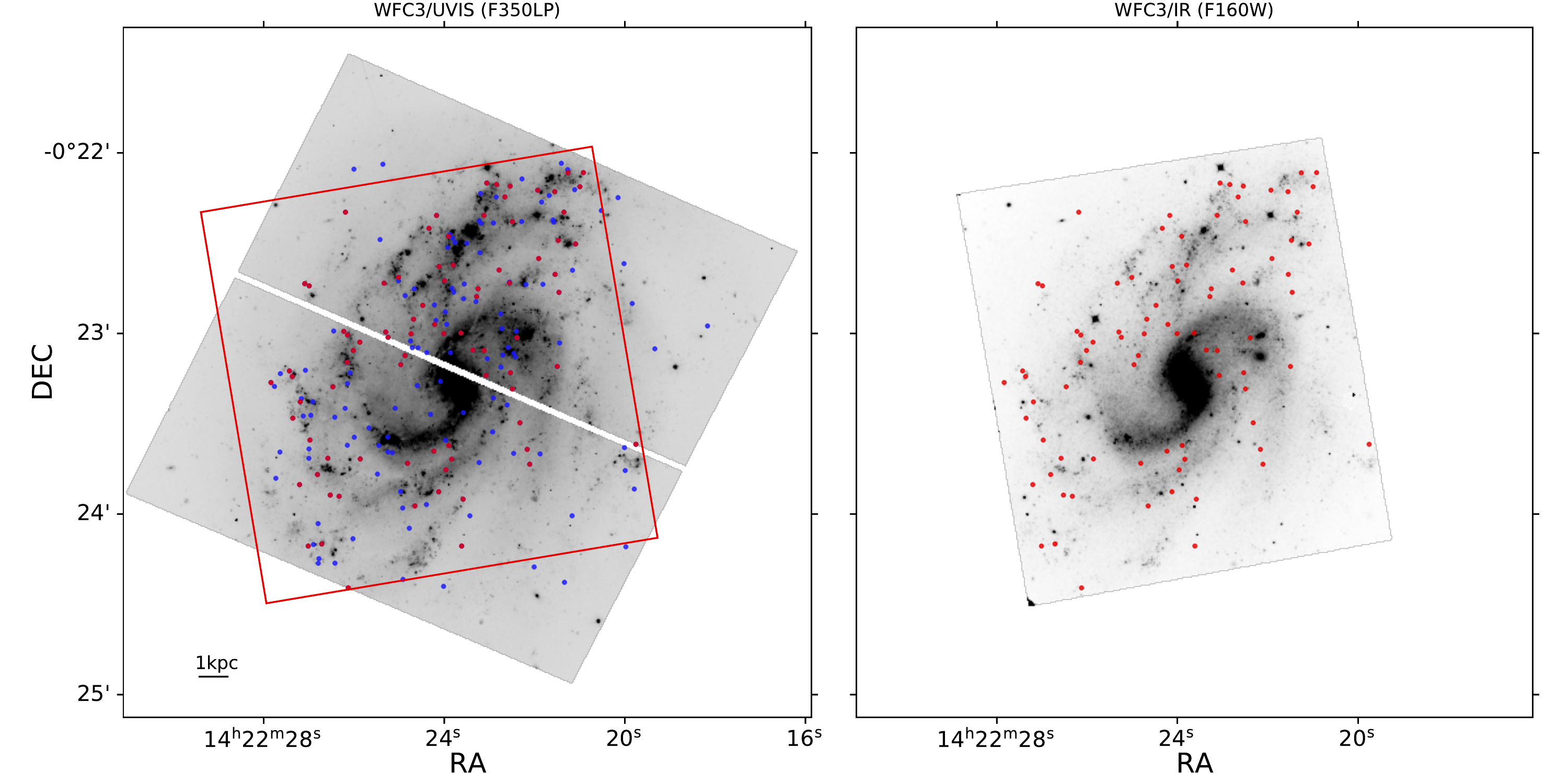}
    \caption{Examples of the HST images from the NGC~5584 in the F350LP (left) and F160W (right) bands. The former image is from the WFC3/UVIS which has two mosaics of 4096 $\times$ 2051 pixels separated by a 35-pixel ($\approx 1.4 \arcsec$) gap, and the latter is from the WFC3/IR with a dimension of 1014 pix $\times$ 1014 pix. The red square outline on the left panel shows the WFC3/IR field of view on this galaxy. The scattered dots show the positions of the Cepheids, the ones with red color are identified in both optical and infrared, while the ones with blue are identified only in the optical bands. On both panels North is up and East is to the left.}. 
    \label{fig:cephs_on_gal}
\end{figure*}

For our inspection, we use the same observations of NGC~5584 that were used by SH0ES. Hence, this work is not a complete reproduction of the original experiment, since we do not repeat the observations themselves. However, where possible, we \textit{intentionally} explore different numerical methods and tools than those used by SH0ES to provide an independent insight to the $H_0$ problem. The goal of this work is to inspect the foundations of the Cepheid distance scale, independently of any input on our analysis from the SH0ES team.

This paper is organized as follows. In Section \ref{sec:method}, we briefly outline the method. Section \ref{sec:data} presents a full description of the data. We describe our analysis in Section \ref{sec:analysis}, present our results in Section \ref{sec:results}, and finally conclude in Section \ref{sec:conclusion}.

\section{Method}\label{sec:method}
A standard approach for distance measurements using the Cepheid variables \citep{Leavitt1912} is to use reddening-free "Wesenheit" index \citep{Madore1982} in the $H$ band defined by \citetalias{Riess2016} as

\begin{equation}\label{eq:wesenheit_mag}
    W_H=H-R_H(V-I)
\end{equation}
where $H$, $V$, and $I$ are mean magnitude of the Cepheids in F160W, F555W, and F814W, respectively. In our analysis, we adopt $R_H=0.386$ which is derived from \citet{Cardelli1989} and \citet{Fitzpatrick1999}, and is also adopted by e.g. \citet{Riess2019ApJ}, \citet{Bentz2019}, and \citet{Breuval2020}. The distance to a nearby SNIa host galaxy can be measured by obtaining the relation between the pulsation period and $W_H$ of its identified Cepheids, and by adopting a $W_H$ vs. period relation calibrated by the Cepheids in e.g. the Milky Way or the LMC. The observations of NGC~5584 for the measurements of periods and the mean magnitudes of its Cepheids in the above-mentioned bands are described in the next section.

\section{Data}\label{sec:data}
\subsection{Archival Observations}\label{sec:archiv_obs}
We obtain the data of NGC~5584 from the Mikulski Archive for Space Telescopes (MAST) database\footnote{\url{https://archive.stsci.edu/access-mast-data}}. NGC~5584 has been observed by the Wide Field Camera 3 (WFC3) between January and April 2010 with the purpose of measuring a Cepheid distance to type Ia supernova SN 2007af hosted by this galaxy (PI: Adam Riess, Cycle: 17, Proposal ID: 11570). WFC3 has been installed on the HST in 2009 replacing the WFPC2. It has two imaging cameras: the UV/Visible channel (UVIS) and the near-infrared (IR) channel. UVIS has two mosaics of 2051 pixel $\times$ 4096 pixel each, a total field of view (FOV) of 162$\times$162 arcsec$^{2}$, and a resolution of 0.04 arcsec/pixel. The IR camera has a dimension of 1014 pix $\times$ 1014 pix with FOV 136$\times$123 arcsec$^{2}$, and a resolution of 0.13 arcsec/pixel.

NGC~5584 has been observed in 13 epochs (in total 45540 sec) in F555W band, 6 of which also accompanied by F814W observations (in total 14400 sec). In twelve of these epochs, this galaxy has also been observed in the F350LP band (in total 15000 sec). In addition, NGC~5584 has also been observed with WFC3/IR channel with the F160W or the H band in 2 epochs (in total 4929 sec). 

\subsection{Calibrations}\label{sec:calibrations}
In all cases, we obtain the calibrated cosmic-ray cleaned (i.e. the \texttt{flc.fits} files for WFC3/UVIS, and \texttt{flt.fits} files for the WFC3/IR observations) data provided by the MAST database. In the case of WFC3/UVIS, the \texttt{flc.fits} files are also corrected for the charge-transfer efficiency loss\footnote{WFC3/IR observations do not suffer from this loss.}. We list the full information regarding these observations in Appendix \ref{appendix_datafiles}.

Similar to \citetalias{Hoffmann2016}, we use the \texttt{TweakReg} software\footnote{\label{drizzlepac}Part of the \texttt{DrizzlePac} software package provided by STScI.} for image registration and alignment. For all the images of all the bands we achieve an alignment better than 0.1 pixels, the same precision is also reported by \citetalias{Hoffmann2016}. We use the coordinates of the "local standard stars" provided by \citetalias{Hoffmann2016} (see section \ref{sec:Cepheids} for more details on these stars) in order to have the same absolute astrometry as theirs. This provides an exact identification of the Cepheids using the RA and DEC reported by \citetalias{Hoffmann2016}.

Each observation epoch consists of multiple exposures. For example, 11 out of 13 epochs in F555W bands consist of six different exposures, and the other two, consist of four different exposures (see Appendix \ref{appendix_datafiles}). We use the \texttt{AstroDrizzle} software$^{\ref{drizzlepac}}$ to combine all the exposures of each epoch (and of the same filter) to obtain final distortion-corrected drizzled science images for the purpose of our analysis. 

Figure \ref{fig:cephs_on_gal} shows examples of UVIS and IR images of the NGC~5584.

\newpage

\subsection{The Cepheids in NGC~5584}\label{sec:Cepheids}
After performing Point-Spread Function (PSF) photometry for all the sources in the galaxy image, \citetalias{Hoffmann2016} uses the \citet{Welch1993} variability index to identify variable objects. This procedure requires comparing fluxes of each epochs with non-variable sources. A list of visually inspected such "local standard stars" is provided by \citetalias{Hoffmann2016} in their table 3. \citetalias{Hoffmann2016} fits all variable objects with Cepheid light curve templates from \citet{Yoachim2009}, which have been generated for V and I bands and by a combination of Fourier decomposition and principal component analysis from a sample of Cepheids in the MW, the LMC, and the Small Magellanic Cloud. After template fitting, \citetalias{Hoffmann2016} visually inspects the six best solution for all the variables, and rejects the variables that are poorly fitted.  The details on further criteria that \citetalias{Hoffmann2016} applied to get to their final Cepheid sample are presented in their section 4.2.

In this work, rather than redoing the Cepheid identification process, we use the same identified Cepheids provided and used by \citetalias{Hoffmann2016} and \citetalias{Riess2016}. This enables us to directly compare our photometry and light curve modeling results with those of SH0ES for each and all of the identified Cepheids in NGC~5584.

\section{Analysis}\label{sec:analysis}
\subsection{Photometry}
Our precise alignment of the images using the local standard stars of \citetalias{Hoffmann2016} provides an exact identification of the Cepheids using the RA and DEC reported by \citetalias{Hoffmann2016}. In the left panel of Figure \ref{fig:cephs_on_gal}, we mark the positions of the 199 optically identified Cepheids. Out of these, only 82 are identified in F160W and measured by \citetalias{Riess2016}. These are marked with red dots on both panels of Figure \ref{fig:cephs_on_gal}.

\begin{figure}
    \centering
    \includegraphics[scale=1]{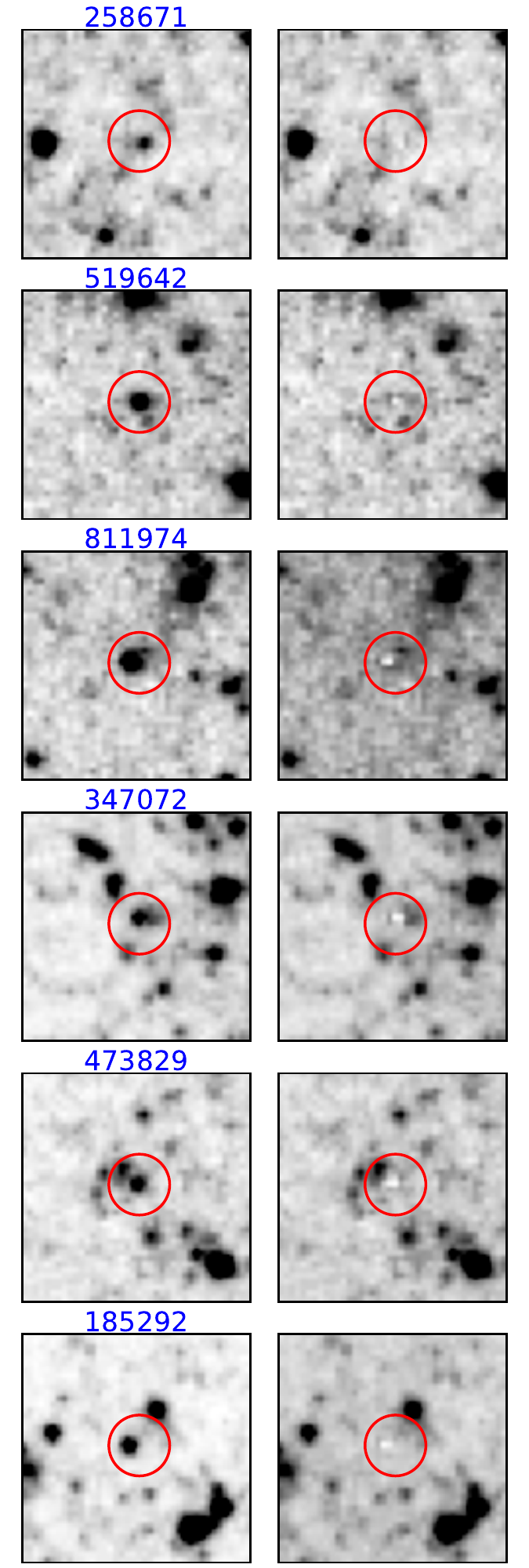}
    \caption{Our PSF photometry for the representative Cepheids in Figure 4 of \citetalias{Hoffmann2016}. The left column shows 40$\times$40 pixels in F555W band centered on each Cepheid with the Cepheid ID (which is the same as in \citetalias{Hoffmann2016}) given on top of each frame. The right column corresponds to the exact same image with the Cepheid removed after the PSF modeling. }
    \label{fig:rep_ceph_psf}
\end{figure}

\subsubsection{PSF Modelling}\label{sec:psf_phot}
To measure the brightness of the Cepheids at each epoch, we use the PSF photometry routines of the \texttt{Photutils} package of \texttt{Astropy} \citep{Bradley2019} that provides tools similar to, but also more general than, DAOPHOT \citep{Stetson1987} which is used by \citetalias{Hoffmann2016}. For optical bands, we perform the PSF photometry on 100 pix $\times$ 100 pix (4 by 4 arcsec$^2$) portions of the image centered on each Cepheid for all the epochs. The background is locally estimated for each Cepheid and is automatically subtracted from the Cepheid flux. Similar to \citetalias{Hoffmann2016}, we use the \texttt{TinyTim} package that provides PSF models for various cameras and different HST bands \citep{Krist2011}. We checked various fitting algorithms and background estimators and found that the choice has negligible effect on the flux measurement. Therefore, similar to \citetalias{Riess2016}, we use a Levenberg–Marquardt-based algorithm (provided by \texttt{Astropy} as \texttt{LevMarLSQFitter}) for determining the best-fit parameters which are the (x,y) position and the flux (plus their uncertainties) for the Cepheids, and \texttt{MMMBackground} routine which calculates the background using the DAOPHOT MMM algorithm \citep{Stetson1987}. 

For IR photometry, i.e. for the F160W band, our procedure is the same as in the optical analysis, except that (similar to \citetalias{Riess2016}) the (x,y) positions of the Cepheids are fixed to their best-fit values from the F814W band and that the PSF photometry is performed on 50 pix $\times$ 50 pix (around 6.5 by 6.5 arcsec$^2$) portions of the image centered on each Cepheid. The reason for fixing the (x,y) position is that the significantly lower resolution of IR images may lead the fitting algorithm to pick a wrong neighbouring source rather than the Cepheids if (x,y) are allowed to vary as free parameters.

Examples of our PSF photometry results are shown in Figure \ref{fig:rep_ceph_psf}. We choose to show those Cepheids that are presented as representative by \citetalias{Hoffmann2016} (in their Figure 4).

\subsubsection{Epoch-to-epoch offset}
The observation condition varies from epoch to epoch and would affect the flux of the Cepheids. To correct for this, we use the local standard stars which were introduced earlier. For each band, we perform a PSF photometry of these non-variable stars and measure their average fluxes in all epochs, $\bar{F}_{\mathrm{all}}$, and also for each epoch, $\bar{F}_{\mathrm{epoch}}$. The Cepheid fluxes at each epoch is then scaled by $\frac{\bar{F}_{\mathrm{all}}}{\bar{F}_{\mathrm{epoch}}}$ to correct for the epoch-to-epoch offset\footnote{Since we scale the Cepheid fluxes by the ratio $\frac{\bar{F}_{\mathrm{all}}}{\bar{F}_{\mathrm{epoch}}}$, the difference between the choice of a statistic (whether mean or median) is negligible.}.

\subsubsection{Magnitude zero-points and aperture correction}\label{sec:zp_apcor}
The magnitude zero-point, ZP, for different HST bands are provided by \citet{Kalirai2009}, \citet{Deustua2017}, and on the STScI calibration pages\footnote{\url{https://www.stsci.edu/hst/instrumentation/wfc3/data-analysis/photometric-calibration}}. These ZP values are based on WFC3 standard aperture radius of 0.4 arcsec. Therefore, the difference between this standard aperture and the PSF modeling should be measured and corrected for. A customary approach adopted also by \citetalias{Hoffmann2016} is to perform both PSF and aperture photometry on a sample of ideally isolated and relatively bright stars in the image and to obtain a statistical mean difference between the two.

In this work, we take a rather different approach. Ideally, for a single isolated star the difference between the aperture and PSF photometry should be directly dependent on the aperture size and the PSF model, while the background should be the same. Here, given that the aperture size for the purpose of correction is fixed to 0.4 arcsec, the difference is basically caused by the extra light captured by the tails of the PSF model beyond 0.4 arcsec radius. Therefore, one way to directly obtain this difference is to measure the flux of the PSF model using a 0.4 arcsec radius aperture (10 pixels for UVIS and around 3 pixels for IR). The magnitude difference, hence the aperture correction ($\mathrm{AP_{cor}}$), would then be
\begin{equation}\label{eq:ap_cor}
\mathrm{AP_{cor}}=2.5\log_{10}[\frac{F_{\mathrm{ap}}}{F_{\mathrm{PSF}}}]-2.5\log_{10}[EE(r=10)], 
\end{equation}
where $F_{\mathrm{ap}}$ is the fraction of the PSF flux inside the aperture, $F_{\mathrm{PSF}}$ is the total flux of the PSF model, and $EE(r)$ is the encircled energy for different aperture radius $r$ \citep[see ][for further details]{Deustua2017}. We compare the two methods of measuring the aperture correction in Appendix \ref{sec:appendix_appcor}.

The ZP and $\mathrm{AP_{cor}}$ values used in this study are listed in Table \ref{tab:zp_apcor} for each band\footnote{We note that R16 and H16 used 25.741, 24.603, and 24.6949 as ZP values for F555W, F814W, and F160W, respectively}.

\begin{table}
    \centering
    \begin{tabular}{c|cccc}
         & F555W & F814W & F350LP & F160W \\
         \hline
        ZP (mag) & 25.737 & 24.598 & 26.708 & 24.5037 \\
        $\mathrm{AP_{cor}}$ (mag) & 0.032 & 0.034 & 0.032 & 0.049\\
    \end{tabular}
    \caption{The zero-point (ZP) and the aperture correction ($\mathrm{AP_{cor}}$) values both in mag for the different HST bands used in this study. See Section \ref{sec:zp_apcor}.}
    \label{tab:zp_apcor}
\end{table}

\subsection{Crowding Bias}\label{sec:crowding}

At distances larger than $\approx10$ Mpc, despite the large luminosity of Cepheids, their light often cannot be separated from their stellar crowds \citep{Riess2020}. The flux of the neighboring stars entering the same resolution element as the Cepheid alters the statistical estimation of the background, therefore biasing the Cepheid flux \citep{Anderson2018}. This bias is one of the most significant challenges for Cepheid measurements at distances larger than 20 Mpc \citep{Freedman2019}. In particular for NGC~5584, at a distance of around 23 Mpc, each pixel of the WFC3/UVIS camera spans around 4 pc. Therefore, it is very likely that the pixel that contains a given Cepheid also encompasses other stellar sources either physically near the Cepheid, or along the line of sight. The pixel size of UVIS/IR is around three times larger than that of WFC3/UVIS, hence covering a larger physical size at the distance mentioned above. The so called "crowding bias" can be statistically estimated at the location of each Cepheid and can be removed from the flux measurements. A typical method, which is also used by the SH0ES team, is to simulate and add artificial stars to the immediate surroundings of each Cepheid on an image, retrieve their flux using the same PSF photometry approach applied to the Cepheids, and measure the difference between the input and output fluxes. In a recent study, \citet{Riess2020} present a test of this approach using an independent method employing the Cepheids amplitudes, and report an statistical agreement between the two methods. 

In this work, we use a similar approach as in \citetalias{Riess2016} and \citetalias{Hoffmann2016}. In the case of the optical bands, for each Cepheid we simulate (using \texttt{TinyTim} PSF models) 20 artificial stars per epoch and add them to the same image portions used for their PSF photometry (Section \ref{sec:psf_phot}). In the case of the F160W band, because only two epochs are available, we use 50 artificial stars per epoch. The fluxes of these artificial stars are then measured using the same PSF approach explained in Section \ref{sec:psf_phot}. Prior to obtaining a mean value for the magnitude differences, \citetalias{Hoffmann2016} directly removes the artificial stars that land within 2.5 pixels of another source that is up to 3.5 mag fainter. Instead of this direct removing approach, we apply a $2\sigma$ clipping which automatically rejects the artificial stars that are blended with another bright source. We then measure the mean magnitude difference as the crowding bias estimate for each Cepheid. 

For the optical observations, the SH0ES team uses the mean value of crowding bias in a galaxy as a single bias value for all the Cepheids in that galaxy. By doing that, the local bias are over estimated for some Cepheids, and are underestimated for some others. Crowding is an environment dependent effect and, in principle, it should not be averaged over a galaxy. In our analysis, we take a different but accurate approach and apply the crowding bias estimated at the position of each Cepheid on the measured magnitudes of that Cepheid before template light curve fitting. We investigate the crowding bias in more detail in Appendix \ref{sec:appendix_crd} where we derive a relation between crowding bias and local surface brightness and we also compare our results with those of SH0ES.

\begin{figure}
    \centering
    \includegraphics[width=0.48\textwidth]{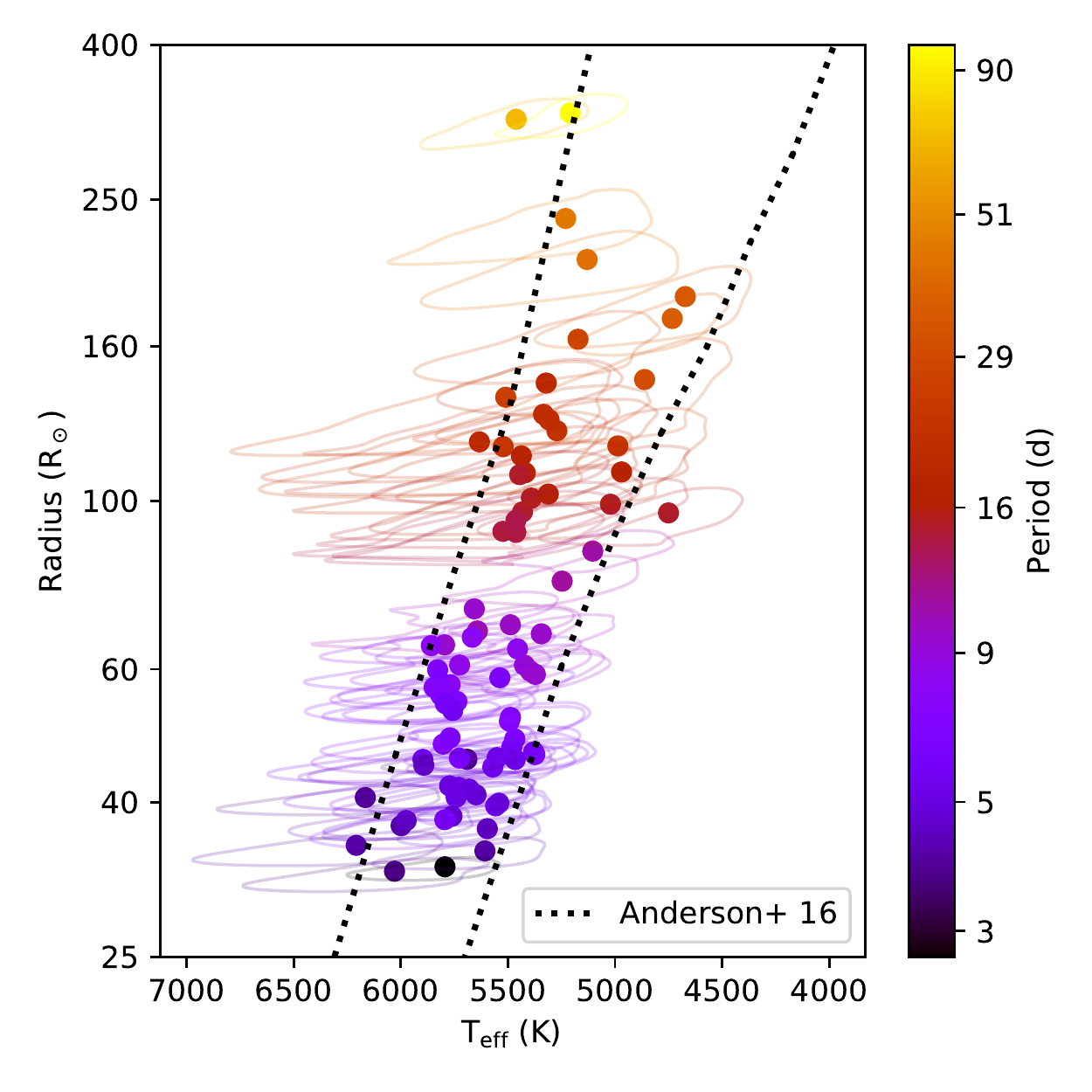}
    \caption{Radius vs. Temperature diagram for the Galactic Cepheids. The dots are the average values (over the pulsation), whereas the thin lines show the values over the pulsation. The color code refers to the pulsation period. The dotted lines are the borders of the theoretical instability strip, using mild rotation (0.5) and solar metallicity \citep{Anderson2016}. We only use the Cepheids with period greater than 12 days for building our template light curves, see Section \ref{sec:spips_data} for further details.}
    \label{fig:template_RT}
\end{figure}

\begin{figure}
    \centering
    \includegraphics[scale=0.68]{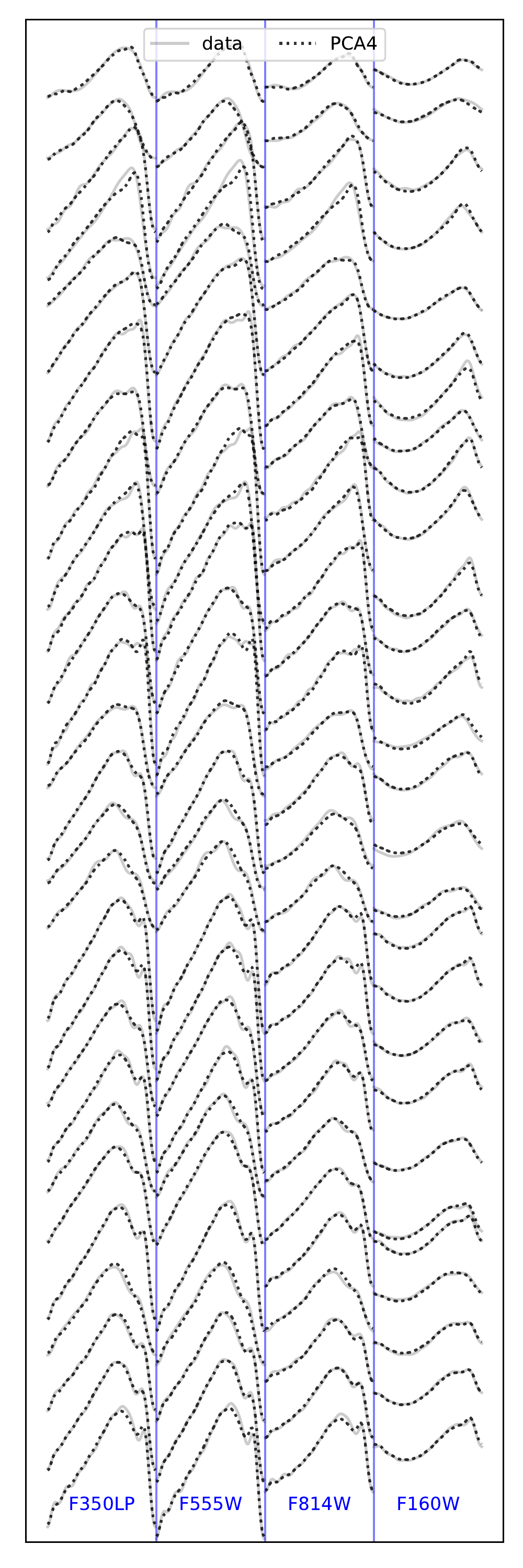}
    \caption{The training set (continuous lines) and reconstructed (doted line) light curves, sorted by pulsation period: from shortest (bottom) to longest (top). See Section \ref{sec:reduction_of_dimensions}.}
    \label{fig:PCA_data_and_model}
\end{figure}

\begin{figure}
    \centering
    \includegraphics[scale=0.7]{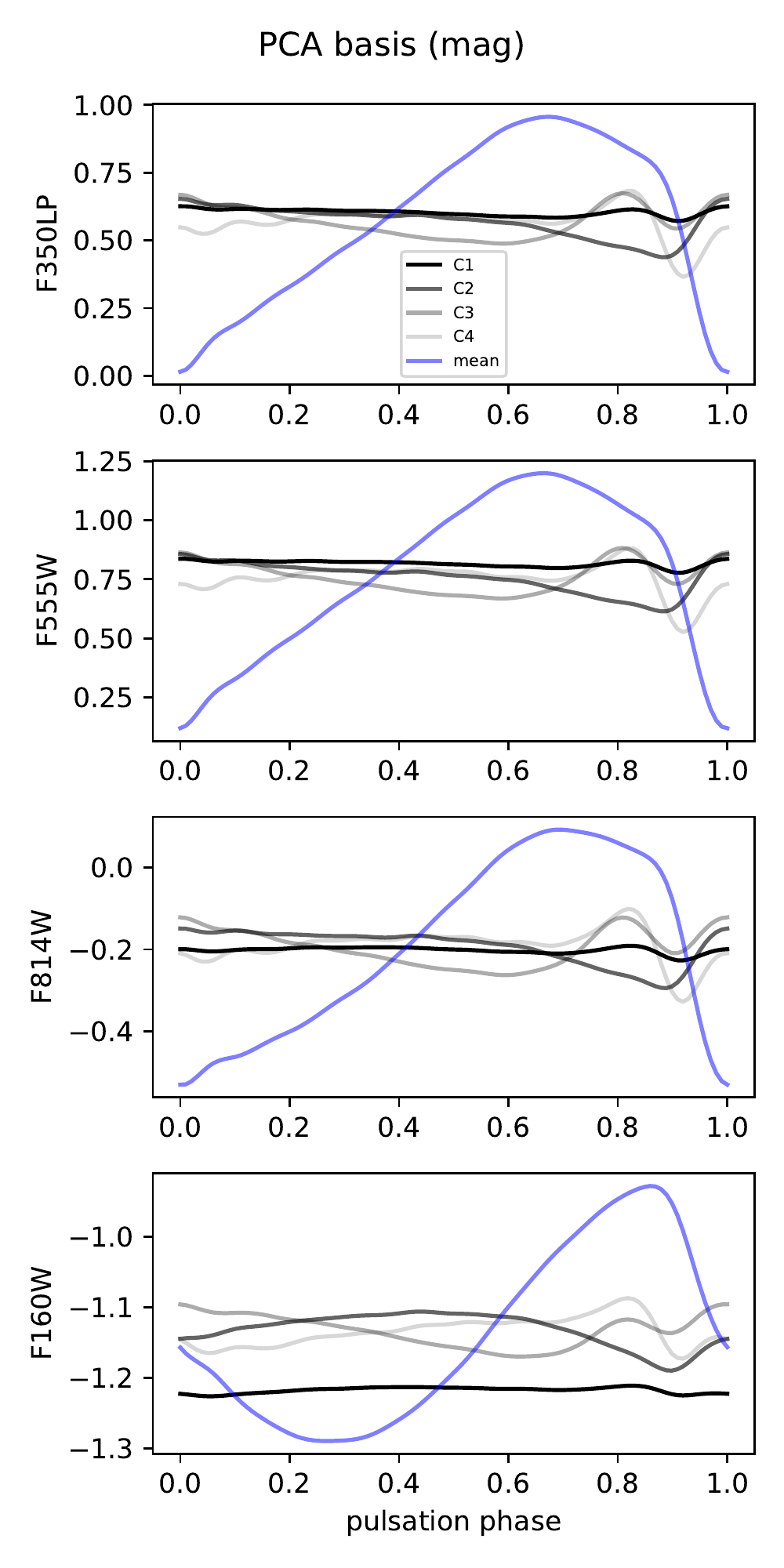}
    \includegraphics[scale=0.65]{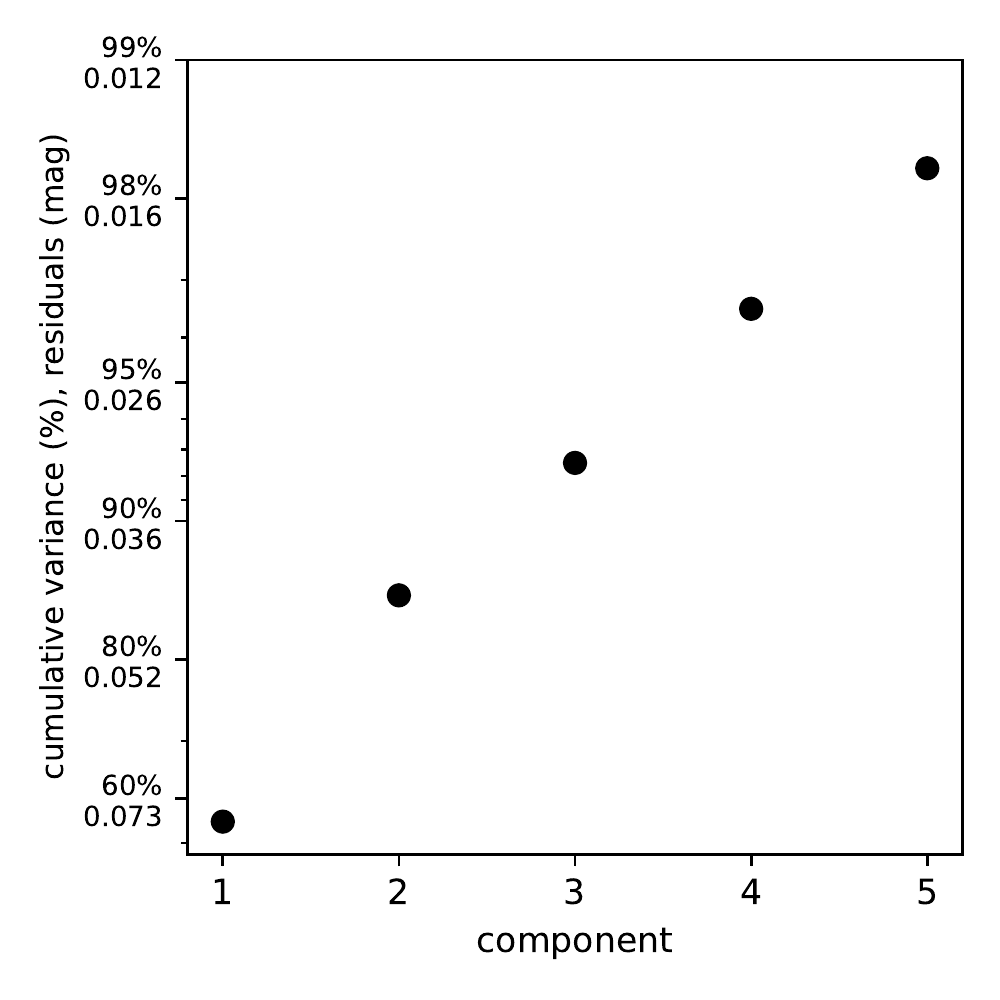}
    \caption{Top four panels: PCA components, in blue the average light curve, and in shades of gray the first 4 components (the darkest is the first component). Bottom: the increase of the training data set variance covered as a function of number of components used in the reconstruction.}
    \label{fig:PCA_components}
\end{figure}

\begin{figure}
    \centering
    \includegraphics[scale=0.6]{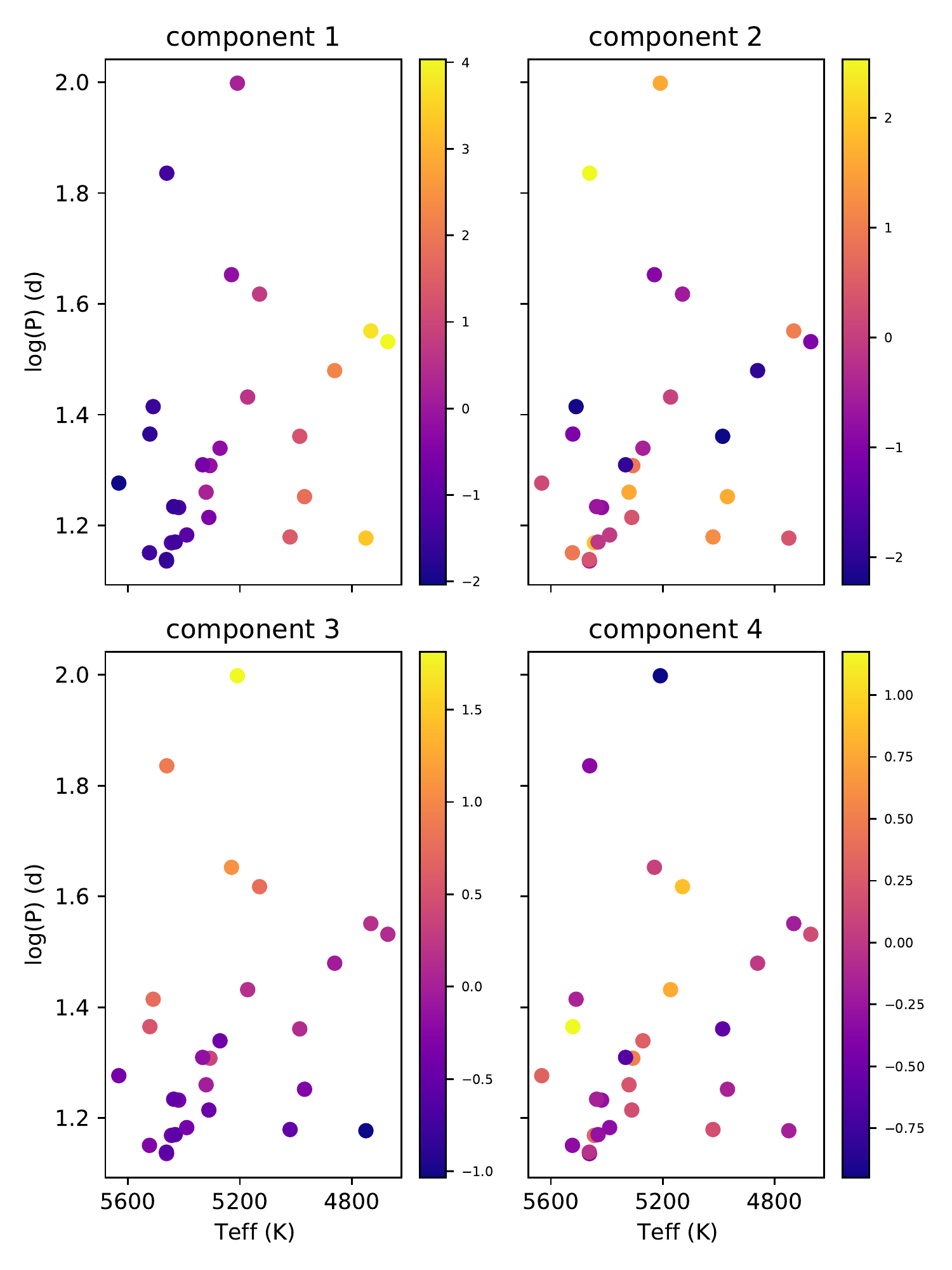}
    \caption{The color coded PCA coefficients as function of Cepheids' effective temperature and period. Interestingly, component 1 is strongly correlated with temperature, and component 3 is strongly correlated with period.}
    \label{fig:PCA_Teff_period}
\end{figure}

\subsection{Light curve fitting using Templates from Galactic Cepheids}\label{sec:templates}

The data collected for each Cepheid consists of several epochs for different pass bands. From this data, we need to derive the pulsation period, as well as the mean magnitudes in each band. In \citetalias{Hoffmann2016}, this was done using template light curves from \citet{Yoachim2009}. In this work, we use different template light curves and fitting strategy so that all bands are analysed simultaneously. 

We derive synthetic light curves in the HST photometric bands for various known Galactic Cepheids, covering the instability strip (in effective temperature and period). The radius, effective temperature, and period of these Cepheids are shown in Figure~\ref{fig:template_RT}. We then use a dimensionality reduction algorithm to parametrize any light curve using only a few parameters.

\subsubsection{Data set for the templates}\label{sec:spips_data}
We choose to use observational data as basis for our templates, fitted with our modeling tool SPIPS \citep{Merand2015} which synthesizes photometric observations based on variations of the stellar radius and effective temperature. We collect high quality spectro-, photo- and interferometric data for many Galactic Cepheids and fit their SPIPS models. It should be noted that the knowledge of the distance and/or the projection factor of these Galactic Cepheid does not play a role in building the light curve templates. 

Our final sample comprises of 28 stars with periods ranging from 12 to $\sim$90 days (Breuval et al., in prep). We do not include Cepheids with period shorter than 12 days because i) Cepheids observed in distant galaxies are biased towards the brightest ones, which results in an observational cut around $\sim$20 days and ii) Cepheids light curves change dramatically around 9-10 days, which has been long noticed ever since Fourier decomposition was applied to Cepheids' light curves \citep[see e.g.][]{Simon1981}. We do not apply any selection cut on radius and effective temperature, as our intent is to sample Cepheids in the instability strip. 

The SPIPS models are based on radial and temperature variations enabling the synthesize of any photometric light curve using the filter band pass definition and atmospheric models. The advantage of this method is that it can accurately extrapolate light curves in pass bands for which we do not have data. We use the HST band passes and zero points defined at the Spanish Virtual Observatory's Filter Profile Service\footnote{\url{http://svo2.cab.inta-csic.es/theory/fps/}}.

\subsubsection{Reduction of dimensions in templates} \label{sec:reduction_of_dimensions}
Our 28 Galactic Cepheids light curves contain a lot of information which needs to be reduced into parametrised templates. We reduce the dimensions of our template data set with a principal component analysis (PCA), using the \texttt{scikit-learn} Python library \citep{scikit-learn}. The training data set for PCA are 28 vectors composed of the concatenated light curves (one for each band) over a single pulsation cycle, centered around their means (see Figure \ref{fig:PCA_data_and_model}). When it comes to choosing how many components to keep to fit our light curves, it is customary to consider the amount of variance reproduced by a given number of most significant components. In our dataset, at any given phase and for any bands, the standard deviation is never greater than 0.2 mag, with a total standard deviation of 0.13 mag (around the average light curve). Using enough PCA components to reproduce 95\% of the variance should reproduce light curves within $\sim$0.03mag (on average), which we deemed enough for our application. Our main goal is to extract the average magnitude from sparse and irregularly sample time series: even if the light curve is reproduced within 0.03 mag, the average is likely estimated with much higher accuracy. Keeping 3 principal components covers 93.8\% of our training set variance, whereas using 4 components leads to 97.1\% of the variance being reproduced, which corresponds to a standard deviation of 0.022 mag. See Figures \ref{fig:PCA_components}, and \ref{fig:PCA_Teff_period} for different information regarding the PCA components.

\subsubsection{Fitting strategy}
For a given NGC~5584 Cepheid, we have a list of observations in various pass bands and at different dates. The initial period is estimated by a computed periodogramme on the F555W and F350LP data. Then, a full model is fitted to the data using the PCA light curves. Our model also includes reddening, using the formula contained in SPIPS and parametrised using the color excess E(B-V). We fix the reddening to E(B-V)=0.035, which we estimate using \texttt{DUST}\footnote{\url{https://irsa.ipac.caltech.edu/applications/DUST/}}

We iterate on the initial parameter by randomising the period ($\pm$5\%) to account for the uncertainty of the periodogramme estimation. The PCA coefficients are initialised to their mean value from the analysis of the template stars. During the least square fit, a uniform prior constrains the coefficient to only evolve inside the range of values observed on the template stars. From the randomised starting periods, we keep the fit with the global lowest reduced $\chi^2$.
Using the best fit parameters and covariance matrix, we can compute the domain of uncertainty for the synthetic light curves and derive the average magnitudes and amplitudes. 

\begin{figure}
    \centering
    \includegraphics[scale=0.58]{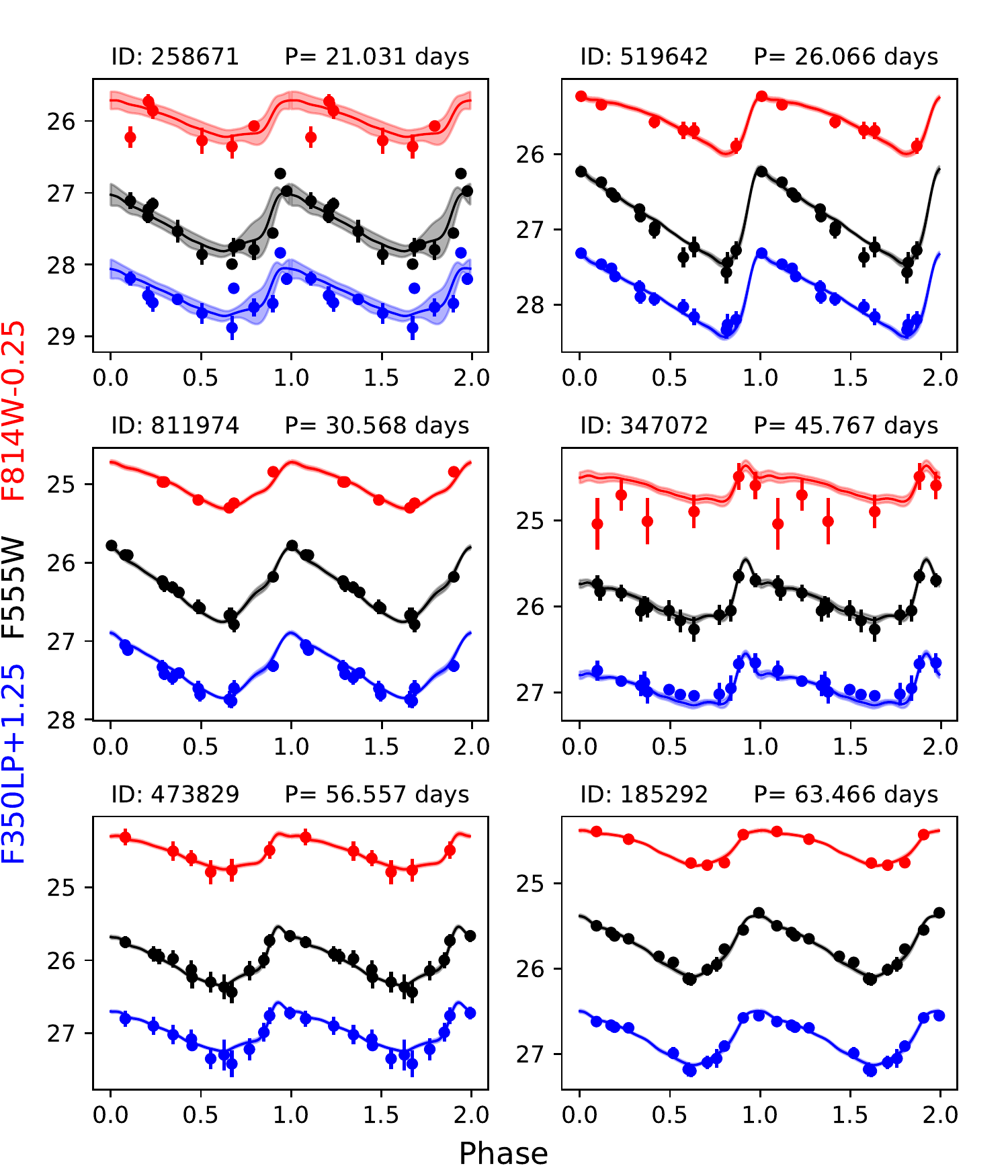}
    \caption{Our light curves of the Cepheids presented by \citetalias{Hoffmann2016} as representative Cepheids (see their Figure 4). On each panel, the bottom (blue), middle (black), and top (red) curves are light curves in F350LP, F555W, and F814W, respectively. Two cycles are plotted and F350LP and F814W have 1.25 and 0.25 mag offsets, respectively. The shaded transparent regions represent the model uncertainties and are present for every curve on all panels. For some curves, they are too small to be seen by eye.}
    \label{fig:rep_LC}
\end{figure}

Our fitting method has several differences with the one presented in \citetalias{Hoffmann2016} using \citet{Yoachim2009}. First of all, we fit all data at once. This is feasible since our model include realistic information about the offset between bands and the shape of the light curve. An example can be seen for star 347072 in Figure~\ref{fig:rep_LC} (which we discuss further in Section \ref{sec:results}). The F814W data of this Cepheid are very noisy and the fitted light curve is constrained mostly by the F555W and F350LP data, which are of much better quality. Even if the modeled light curve in F814W is systematically above the data points, it is the most realistic within our hypothesis and priors derived from Galactic Cepheids. 

\begin{figure*}
    \centering
    \includegraphics[width=\textwidth]{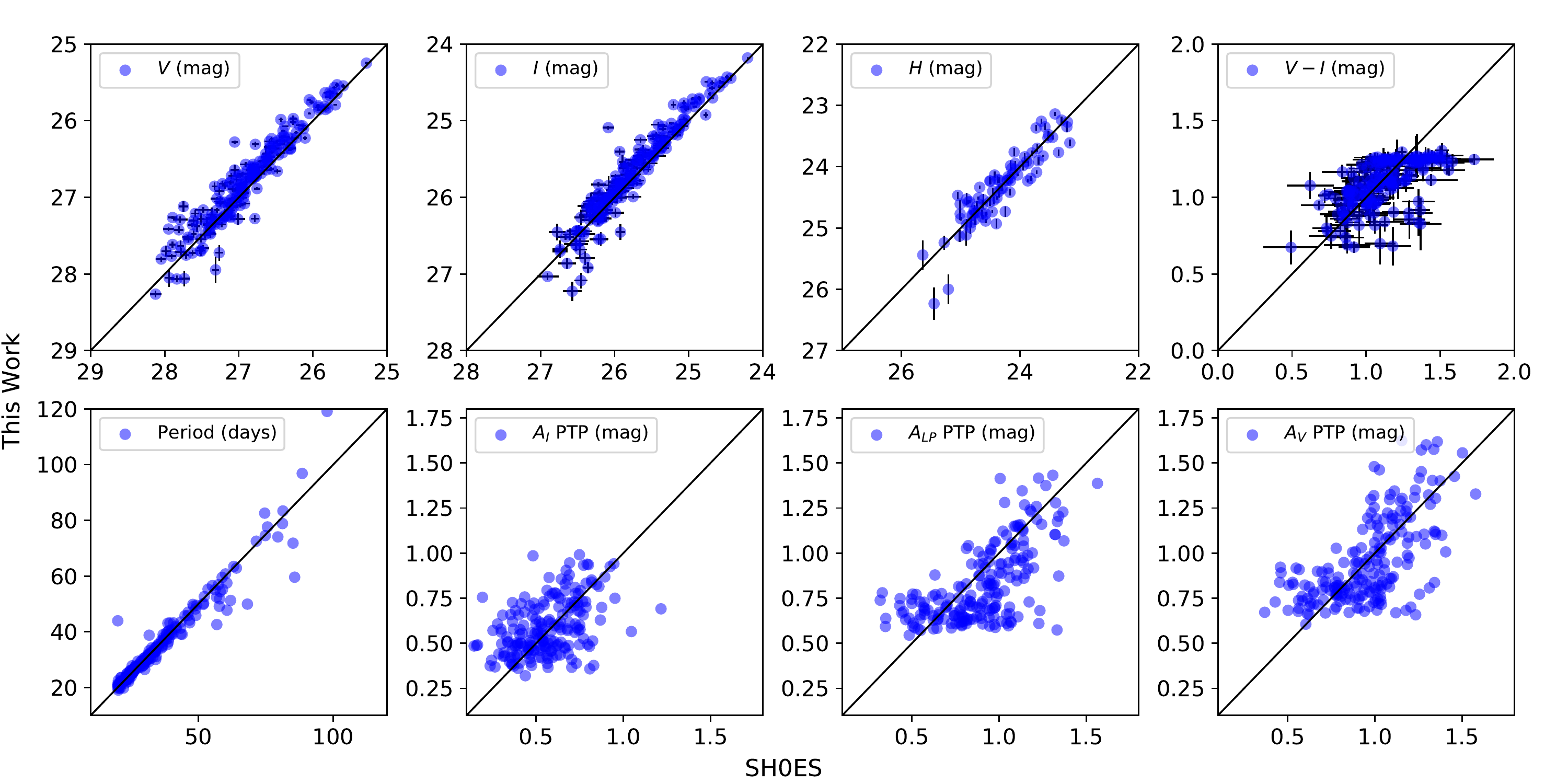}
    \caption{Comparing our results (Y axes) with those the SH0ES (X axes) team for mean magnitudes (LP: F350LP, V: F555W, I: F814W, H:F160W), $V-I$ color, period, and light curve amplitudes ($A$) of the Cepheids in NGC~5584. For the H band, we only show the uncertainties on the Y axis (i.e. from our results), since \citetalias{Riess2016} publishes only the so called total uncertainties ($\sigma_{tot}$) and not those of the mean magnitudes in the $H$ band. PTP stands for peak-to-peak and is one of the methods of determining the pulsation amplitudes (see Section \ref{sec:amplitudes} for more details). The equality lines are plotted with solid black on all panels.}
    \label{fig:comparison_plots}
\end{figure*}

\section{Results}\label{sec:results}
\subsection{Light Curves, Mean Magnitudes, and Periods}\label{sec:mag_and_per}
Using the light curve template fitting explained in Section \ref{sec:templates}, we obtain the periods and the mean magnitudes for all the identified Cepheids in the four HST bands. Figure \ref{fig:rep_LC} presents our results for the light curves of the Cepheids we showed in Figure \ref{fig:rep_ceph_psf}, they are chosen by \citetalias{Hoffmann2016} as the representative Cepheids of NGC~5584. Our light curves can be directly compared with those of \citetalias{Hoffmann2016} shown in their Figure 4. Most of the light curve models nicely represent the data. One exception among these is the Cepheid 347072, which as discussed earlier, has poor quality data points in F814W. This Cepheid is not detected in the F160W band and therefore is not included in the measurement of distance neither by SH0ES nor by us in this work. 

Figure \ref{fig:comparison_plots} provides one-on-one comparisons of our results with those of SH0ES reported in \citetalias{Hoffmann2016} and \citetalias{Riess2016}. The top row provides comparisons for the mean magnitude measurements in $V$, $I$, and $H$ bands, as well as the $(V-I)$ color. For the H band, we only show the uncertainties on the Y axis (i.e. from our results), since \citetalias{Riess2016} publishes only the so called total uncertainties ($\sigma_{tot}$) and not those of the mean magnitudes in the $H$ band. A generally good agreement can be seen between the two results, especially for the H band and the $V-I$ color both of which directly contribute to the distance measurement (see Equation \ref{eq:wesenheit_mag}). 

The left-most panel on the bottom row of Figure \ref{fig:comparison_plots} provides a comparison for period measurements. As can be seen, although we use a different approach for template fitting and hence the period measurements, the two results are in general agreement with only a few exceptions. 

Regrettably, \citetalias{Hoffmann2016} does not provide mean magnitudes in F350LP band, we therefore cannot have a direct comparison for this quantity. However, we can compare amplitude measurements in F350LP band as discussed in the next sub-section.

\begin{figure}
    \centering
    \includegraphics[scale=0.65]{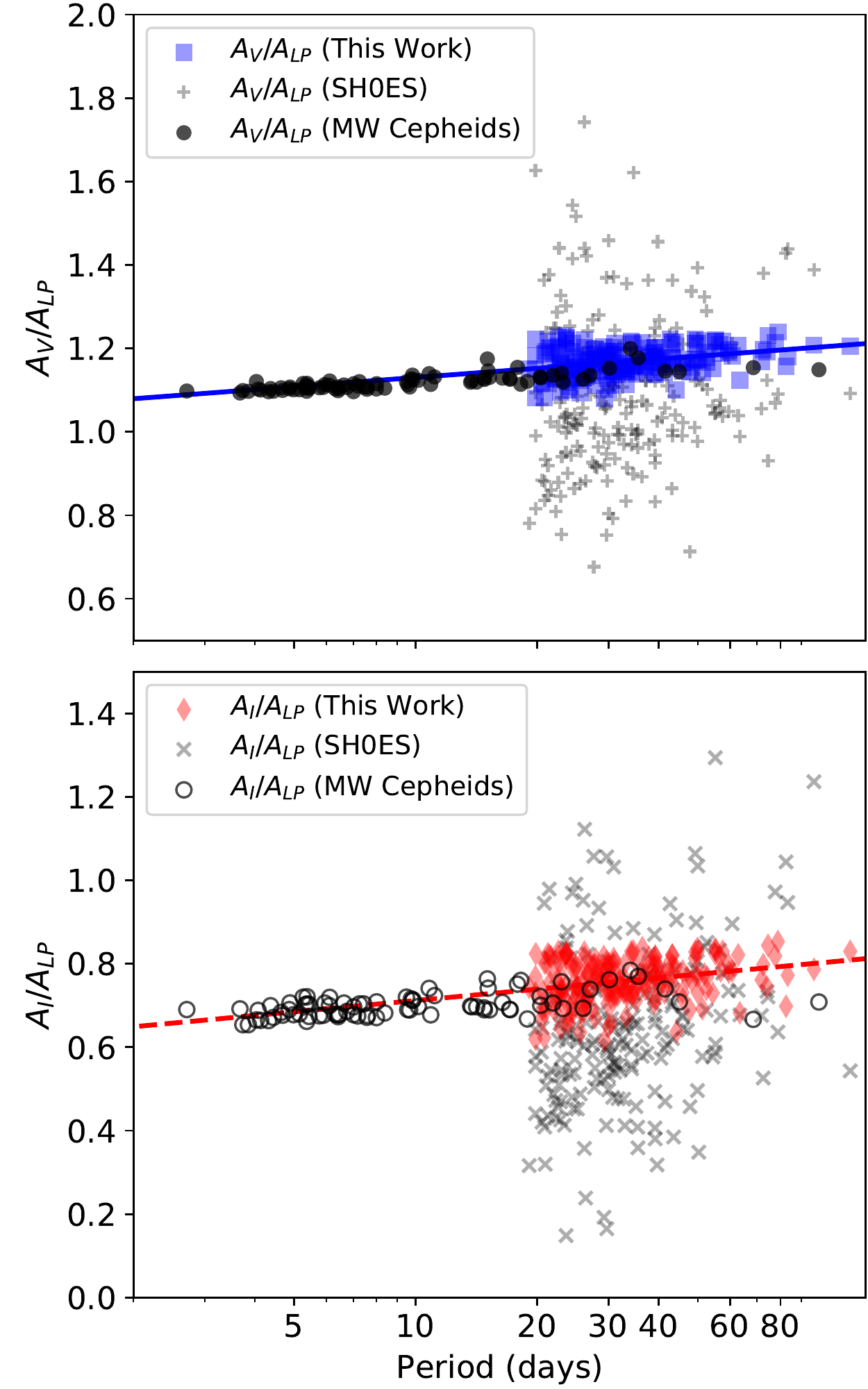}
    \caption{RMS (root-mean-square) amplitudes in V (top panel) and I (bottom panel) bands relative to F350LP (LP) band vs. period. The plus and cross symbols are SH0ES results for $A_V/A_{LP}$ and $A_I/A_{LP}$, respectively. Our results have a significantly lower scatter than those of SH0ES. The solid blue and red dashed lines are linear fits as explained in Section \ref{sec:amplitudes} and shown in Equation \ref{eq:amp_ratio_fit}. We also show the MW Cepheids as filled dots and empty circles, for $A_V/A_{LP}$ and $A_I/A_{LP}$, respectively. We note that while for the linear fitting only the results from the Cepheids in NGC~5584 were used, and while only 28 of the MW Cepheids with period$>$12 days were used for our template light curve building, the fitted line passes also through the Milky Way Cepheids data points.}
    \label{fig:amp_ratios}
\end{figure}

\subsection{Amplitude Measurements}\label{sec:amplitudes}
We remind the reader that \citetalias{Hoffmann2016} uses the amplitude ratios vs. period relation of the Cepheids in NGC~5584 to correct the random-phase observations of other SNIa host galaxies in the $V$ and $I$ band. Therefore, an accurate and precise measurement of these relations can potentially impact the final $H_0$ measurements. The three panels (from the right) on the bottom row of Figure \ref{fig:comparison_plots} provide comparisons for our amplitude measurements vs. those of the SH0ES team. We perform two different measurements of the amplitudes: 1) peak-to-peak (PTP) which measures the magnitude difference between the maximum and minimum of the light curve model, and 2) the root-mean-square (RMS) which is the standard deviation of the light curve (regularly sampled) from their mean value. While the PTP results (which are the ones shown in Figure \ref{fig:comparison_plots}) are in general agreement with the amplitude measurements of SH0ES, it is not robustly estimated in our method. Our PCA-based fits allow variations in the shape of the model, especially between phase 0.8 an 1.0, which is where the amplitude is measured (see for example F350LP light curve of star 258671 in Figure \ref{fig:rep_LC}). On the other hand, the amplitude is directly one of the template fitting parameter in the SH0ES analysis. While PTP and RMS differ by a factor of $2\sqrt{2}\approx2.83$ for a pure sinusoidal wave, the value varies with the exact shape of the light curve. From our high definition template sample star, we find that $\left(\frac{\mathrm{PTP}}{\mathrm{RMS}}\right)_{I}=3.08\pm0.12$ and $\left(\frac{\mathrm{PTP}}{\mathrm{RMS}}\right)_{V}=3.14\pm0.14$. We are interested in ratios between bands, and the comparison with SH0ES' results. For the amplitude ratios we find that $\left(\frac{\mathrm{PTP}_I}{\mathrm{PTP}_V}\right) = 0.98\pm0.03 \left(\frac{\mathrm{RMS}_I}{\mathrm{RMS}_V}\right)$. In other words, the ratio of amplitudes are almost independent of the amplitude measurement method and our amplitude ratios computed from RMS (which we use in our subsequent analysis) are comparable to those of SH0ES with a scatter of 6\%.

In Figure \ref{fig:amp_ratios}, we compare our results and those of SH0ES for the amplitude ratio vs. period relation. The blue squares and red diamonds are our $A_V/A_{LP}$ and $A_I/A_{LP}$, respectively. The grey plus and cross symbols are the same quantities as published by SH0ES in \citetalias{Hoffmann2016} and have significantly larger scatters. The dots and empty circles are, respectively, $A_V/A_{LP}$ and $A_I/A_{LP}$ for the Milky Way Cepheids. The blue solid line, and the red dashed line are our results of linear fits on $A_V/A_{LP}$ and $A_I/A_{LP}$ vs. $\log P$. While for the linear fitting we only used the data from the Cepheids in NGC~5584, and while only 28 of the MW Cepheids with period$>$12 days were used for our template light curve building, the fitted line passes also through the Milky Way Cepheids data points even for those with small periods. The linear correlation coefficient measured for both of these relations are $\approx 0.3$. From the linear fitting we find

\begin{equation}\label{eq:amp_ratio_fit}
\begin{split}
    & A_V/A_{LP}=1.167+0.073(\log P-1.5), \: \sigma_{fit}= 0.014,\\
    & A_I/A_{LP}=0.757+0.090(\log P-1.5), \: \sigma_{fit}= 0.022,
\end{split}    
\end{equation}
where $\sigma_{fit}$ values are the standard deviations of the fits and are an order of magnitude smaller than those of SH0ES (see Table 2 of \citetalias{Hoffmann2016}). The small scatter in this relation means that our amplitude ratio measurement is less noisy and is indicative of a high quality light curve modelling approach. We note that in \citetalias{Hoffmann2016} the light curves of different bands are fitted separately (using \citet{Yoachim2009} templates), and then the amplitudes resulting from the different fits are divided to yield the amplitude ratios. This could be the reason for the large scatter in their amplitude ratios. On the other hand, in our approach, all the light curves (of all bands) are fitted simultaneously, hence the amplitudes are not estimated independently from one another, leading to a lower scatter.

\subsection{Uncertainties on the Wesenheit H Magnitudes}\label{sec:uncertainties}
In their section 2.2, \citetalias{Riess2016} describe a $\sigma_{tot}$ as the total uncertainty on their Cepheid distance measurements. They refer to the uncertainty of the crowding bias in the H band as $\sigma_{sky}$ and that of the optical observations as $\sigma_{ct}$ and they add them as a single value for all the Cepheids in a given galaxy. Since we apply the crowding bias (in all bands) for each Cepheid before the template fitting, the values of mean magnitudes already include the effect of crowding bias and their uncertainties. In addition, our template light curve fitting method analyses all the data together, therefore, the uncertainty on the H band mean magnitudes, already includes the effect of limited phase coverage.

Therefore, for the total uncertainty on $W_H$ we have
\begin{equation}\label{eq:wesenheit_err}
    \sigma_{WH}=[\sigma^2_H+R_H^2(\sigma_V^2+\sigma_I^2)+\sigma^2_{int}]^{1/2},
\end{equation}
where $\sigma_{int}$ is the intrinsic dispersion due to the nonzero width of the instability strip. To estimate $\sigma_{int}$, we follow the procedure of \citet{Riess2019ApJ}. Using the Cepheid observations in the LMC, \citet{Riess2019ApJ} present PL relations and their scatter in different HST bands. To estimate $\sigma_{int}$, they subtract (in quadrature) the mean Cepheid measurement errors from the scatter of the PL relation for a given band. Their mean measurement error for different bands are given in their section 2.2 and the values for the scatter of the PL relations are listed in their Table 3. For $W_H$, the intrinsic dispersion is estimated to be $\sigma_{int}=0.069$ mag.

\begin{figure*}
    \centering
    \includegraphics[width=\textwidth]{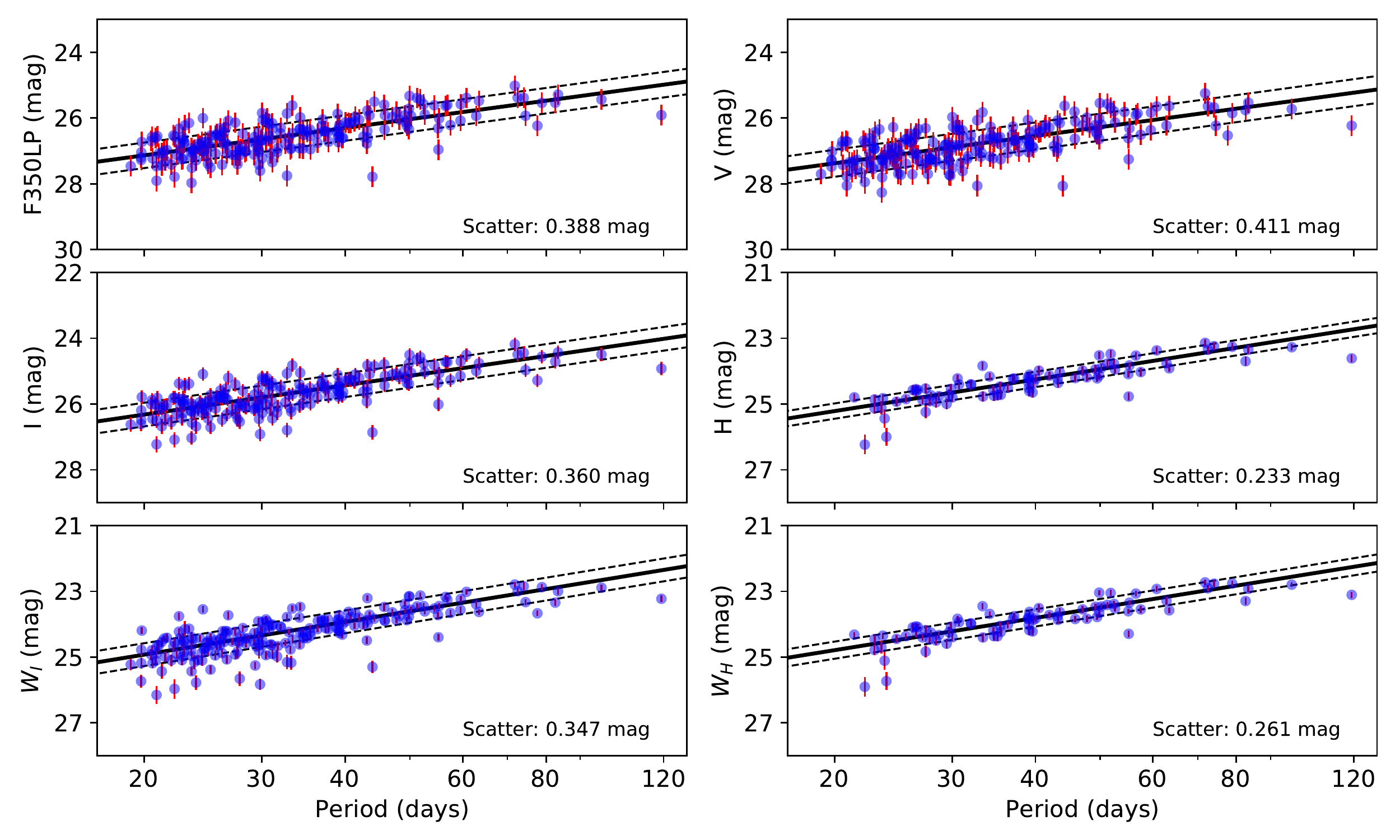}
    \caption{Period-Luminosity relations in the four HST bands and also for the $W_I$ and $W_H$ Wesenheit indices. The uncertainties on individual Cepheids in this figure also includes the contribution from the $\sigma_{int}$ explained in Section \ref{sec:uncertainties}. The solid lines represent the results of fitting a linear relation of the form $m=\alpha\log P+\beta$ with a $3\sigma$ clipping. The slopes ($\alpha)$ are fixed to the values for LMC Cepheids given in Table 3 of \citet{Riess2019ApJ}, see Section \ref{sec:PL_relation}.}
    \label{fig:PL_plots}
\end{figure*}

\subsection{The Period-Luminosity Relations}\label{sec:PL_relation}
In addition to the PL relation in $W_H$, we also present PL relations for all the bands F350LP, F555W, F814W, F160W, as well as for optical Wesenheit index, $W_I$ in Figure \ref{fig:PL_plots}. The latter is defined as $W_I=I-R_I(V-I)$ with $R_I=1.3$ \citep{Riess2019ApJ}. The uncertainties on individual Cepheids in this figure also includes the contribution from the $\sigma_{int}$ explained in the previous section\footnote{We calculate the $\sigma_{int}$ for different bands based on the information given in section 2.2 and Table 3 of \citet{Riess2019ApJ} in the same way as explained in Section \ref{sec:uncertainties}.}. We note that the data points in the PL relations shown in Figure 6 of \citetalias{Hoffmann2016} appear to contain only the measurement uncertainties which are comparable in size to this work's results as shown in our Figure \ref{fig:comparison_plots}. 

The solid lines represent the results of fitting a linear relation of the form $m=\alpha\log P+\beta$, where $m$ is the mean magnitude. We fix the slope $\alpha$ to the values given in Table 3 of \citet{Riess2019ApJ} (which lists the PL relations from \citet{Soszynski2008}, \citet{Macri2015}, and \citetalias{Riess2016}), and fit for the intercept with a $3\sigma$ clipping. The slightly larger scatter in our PL relations compared to those found by SH0ES for NGC~5584 is most probably due to our different treatment of the crowding bias. As stated earlier in the text, SH0ES add a single value of crowding bias for all the Cepheids in a galaxy which shifts the PL relation slightly towards fainter values. However, we add the crowding bias values estimated at the location of each Cepheid separately which introduces a somewhat larger scatter in the PL relation\footnote{We note that the scatter in the PL relation is not influenced by amplitude ratios which together with mean magnitudes are both products of the same template fitting.}.

\subsection{The Distance to NGC~5584}
In this section, we derive the distance modulus of NGC~5584, based on apparent Wesenheit magnitudes $W_H$ of our sample of 82 Cepheids in this galaxy. By applying an existing $W_H$ PL relation to the known period of our stars, we derive the absolute magnitude $M^W_H$ for each Cepheid and then their individual distance modulus $\mu = W_H - M^W_H$. 

\begin{figure}
    \centering
    \includegraphics[scale=0.55]{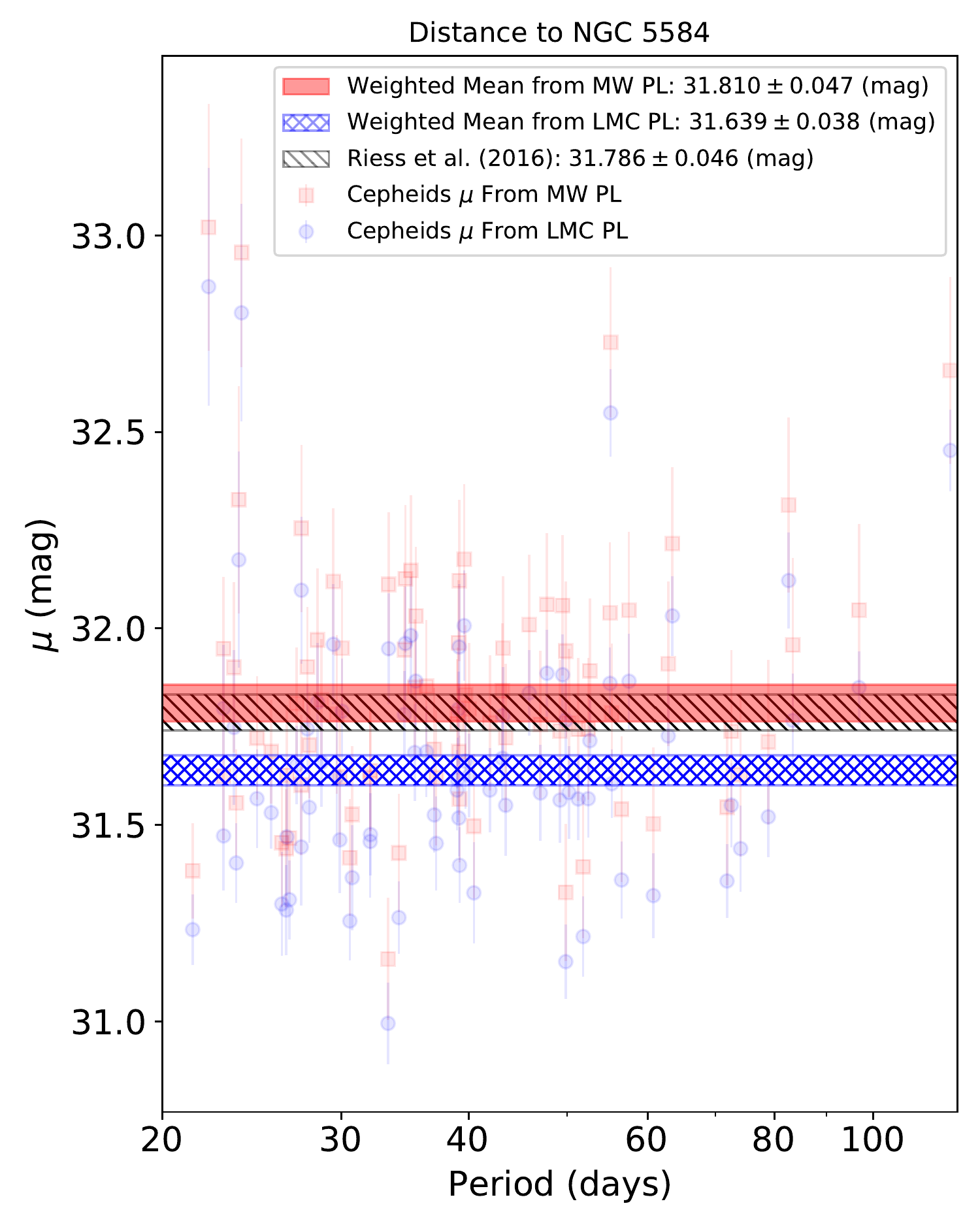}
    \caption{Distance modulus $\mu$ measured from $W_H$ vs. period for Cepheids in NGC~5584 using two PL relations from MW (red squares) and LMC (blue circles) Cepheids. The horizontal filled rectangles show the $1\sigma$ confidence regions for measured distances. The uniformly red and the blue crossed-diagonal hatched regions represent our measurements based on the MW and LMC PL relations, respectively, and are the weighted means of the $\mu$ values measured for individual identified Cepheids. The black back-diagonal hatched region represent the estimated distance reported by \citetalias{Riess2016}. }
    \label{fig:distance}
\end{figure}

We perform this calculation using two different PL relations: one calibrated in the Milky Way \citep{Breuval2020}, $M^W_H = -5.432\:(\pm 0.029) -3.332\:(\pm 0.177) [\log P -0.84]$, and another calibration from the LMC \citep{Riess2019ApJ}, $M^W_H = 15.898 -3.26 \log P$. For the slope of the latter relation, a 0.02 mag uncertainty is stated in \citet{Riess2019ApJ} while they mention no uncertainty on the intercept. Therefore, we assume a conservative uncertainty of 0.02 mag error also for the intercept (the intercept uncertainties in \citet{Macri2015} are much smaller than 0.02 mag). We then subtract the LMC distance modulus as measured by \citet{Pietrzynski2019Nat}. For both PL relations, the individual distance moduli obtained for each Cepheid are represented in Figure \ref{fig:distance}. The Galactic PL relation yields a weighted mean distance modulus of $31.810 \pm 0.047$ mag, while the LMC calibration results in $31.639 \pm 0.038$ mag. The $1\sigma$ confidence regions of these weighted mean values are also shown in Figure \ref{fig:distance}. The distance modulus from the Galactic PL relation is in agreement with $\mu=31.786\pm0.046$ (mag) measured by SH0ES in \citetalias{Riess2016}. The distance modulus from the LMC PL relation, however, is smaller though still in agreement within $2.5\sigma$ with SH0ES result. 

It is not surprising to obtain different distances based on LMC and MW PL relations, given that LMC has a smaller metallicity compared to the MW \citep{Romaniello2008}, i.e. the larger distance inferred from LMC PL relation is consistent with its smaller metallicity. The difference in terms of distance modulus obtained with MW and LMC PL relations highlights the need for a metallicity correction which has been extensively studied \citep[see e.g.][and Breuval et al.~2021, in prep.]{Pietrzynski2004,Gieren2018, Groenewegen2018, Ripepi2019, Ripepi2020}, though yet with no clear consensus. However, NGC~5584 is a spiral galaxy with a structure similar to that of MW and, in fact, its metallicity gradient is very similar to that of the MW \citep[see table 6 of][and table 8 of \citetalias{Riess2016}]{Balser2011}. The MW PL relation, therefore, is more appropriate for measuring the distance to NGC~5584.

From our $\mu$ for NGC~5584 based on MW PL relation, we calculate a distance of $d_{\mathrm{NGC~5584}}=23.01\pm0.05$ Mpc.

\section{Conclusion}\label{sec:conclusion}
The $4-6\sigma$ tension \citep{Riess2019Nat} between the direct and the early Universe measurements of $H_0$ asks for detailed investigations in the different methods involved. NGC~5584 played a key role in the direct measurement of $H_0$ from the Cepheid distance ladder by the SH0ES team \citep{Riess2016}. Observations of this galaxy was employed to derive a relation between the ratio of pulsation amplitude of Cepheids in $V$ and $I$ bands relative to the wide F350LP HST band and the period. The F350LP band has been used by the SH0ES team for detection and light curve measurement of Cepheids in around half of the current SNIa host galaxies used for $H_0$ measurements and the relation mentioned above has been used to obtain mean $V$ and $I$ magnitudes from the spars sampling of Cepheid light curves in these bands.

In this contribution, we provided an independent and detailed analysis of the HST data from NGC~5584. Where possible, we intentionally used methods and tools different from those used by SH0ES. This allowed the investigation of possible influence of these methods on distance measurements. The key parts of our detailed analysis can be listed as follows:
\begin{itemize}
    \item applying PSF photometry routines of \texttt{Photutils} package of \texttt{Astropy} \citep{Bradley2019}, instead of the classic DAOPHOT software \citep{Stetson1987},
    
    \item testing and finding negligible influence of the choice of PSF modelling and background subtraction algorithms,
    
    \item applying a different aperture correction procedure for the PSF photometry, 
    
    \item adopting a slightly modified approach for crowding bias estimation (using a sigma-clipping approach on the artificial stars flux measurement rather than directly removing bright estimated sources done by SH0ES),
    
    \item a different approach for applying the crowding bias compared to SH0ES (applying the bias separately for each Cepheid rather than averaging over the whole galaxy for the optical observations), and
    
    \item employing a completely different approach for Cepheid light curve modelling for measurement of mean magnitudes, amplitudes, and periods.
    
\end{itemize}

And our main results can be summarised as follows:

\begin{itemize}
    \item Our measurements of Cepheids' mean magnitudes and period and those of SH0ES are in good agreement. In particular, we find no systematic difference in our H band mean magnitudes and (V-I) color, both of which directly influence the distance measurements, compared to SH0ES.
    
    \item We derived a significantly tighter amplitude ratio vs. period relation compared to the one derived by SH0ES.
    
    \item We measure two distance moduli for NGC~5584 using two different PL relations calibrated in MW and LMC. The result from the former is in agreement with the value from SHOES within $1\sigma$, and the result from the latter is $0.147\pm0.060$ mag smaller than that of SH0ES, though still within $2.5\sigma$.
\end{itemize}
We do not attempt at reporting a value for $H_0$ based on the distance to only one SNIa host galaxy, and we only note that a smaller distance to NGC~5584 points towards a higher $H_0$ value. However, we consider the MW PL relation to be more appropriate for distance measurements to NGC~5584, due to similar metallicity and structure of these two galaxies. Nevertheless, the effect of metallicity and its measurement methods \citep{Bresolin2009,Kudritzki2012ApJ} on extragalactic Cepheid distances requires further investigations.

The main conclusion of the current study is that our inspection of NGC~5584 Cepheids does not yield any systematic hints towards the resolution of the $H_0$ problem. However, it would be important to also independently inspect for systematics in the distance measurements to all the galaxies used for calibration of SNeIa absolute magnitude. For doing so, and until reasonably fine-sampled time series data of all SNeIa calibrators become available, it would certainly be better to use our precise amplitude ratio vs. period relations for light curve analysis of Cepheids in SNeIa hosts with limited time series data as they would potentially yield more accurate mean magnitudes in V and I bands. This would also provide an investigation into the potential statistical effect of these relations in $H_0$ measurements.

While it is important to continue the investigations on the $H_0$ measurements, the current findings seems to be pointing towards a non-trivial solution to this problem. This could mean that our current understanding of the local or early Universe may require modifications or a complete change of paradigm. In the local Universe, presence of a large local underdensity \citep[which is incompatible with LCDM,][]{Haslbauer2020} has been presented \citep{Shanks2019,Haslbauer2020} as a possible cause of the $H_0$ discrepancy (but see also \citeauthor{Riess2018} \citeyear{Riess2018} and \citeauthor{Shanks2018} \citeyear{Shanks2018}). In the early Universe, various scenarios such as non-standard recombination, dark matter/dark energy interaction, and self-interacting neutrinos have been presented, however, so far no consensus has been reached \citep[for reviews and summaries see][]{Verde2019Nat,Poulin2020,Knox2020}.

While it is important to seek alternative ideas on the theoretical side, the improvement of current observational methods as well as the development of new independent ones are necessary for a progress towards a solution to the $H_0$ problem. For the Cepheid distance ladder, the number of SNeIa calibrators observed by the HST is soon to be doubled by the SH0ES program \citep{Riess2019ApJ}, hence the statistical uncertainty on $H_0$ measured by this method would be reduced. In addition to strong lensing, megamasers, and TRGB methods \citep[see also][]{Beaton2016,Kim2020} mentioned in the Introduction, other Cepheid-independent routes would also soon contribute to $H_0$ measurements. \citet{Huang2020} presents Mira variables for calibration of SNeIa absolute magnitudes. Also, using the advanced LIGO and Virgo gravitational wave detectors, \citet{LIGO-VIRGO-2019} have reported an $H_0$ measurement using standard sirens \citep[see also][]{Coughlin2020}. As the number of detected standard sirens increases in future, the currently large statistical uncertainty in their resulting $H_0$ measurement would decrease, making them an important independent way of measuring the cosmic expansion rate \citep{Feeney2019}. 

One of the most promising contributions to the accuracy of the cosmic distance scale in the near future would be from Gaia. The impact of the first \citep[see, e.g.,][]{2017A&A...599A..67C,2017A&A...605A..79G} and second \citep[see, e.g.,][]{Groenewegen2018,2018ApJ...861..126R,2019A&A...622A..60C,Breuval2020,Ripepi2020} data releases of Gaia on the calibration of the Cepheid PL relation is already considerable. It is however still limited by the persistently uncertain value of the instrumental parallax zero point \citep[see, e.g.,][]{2018A&A...616A..17A,2019A&A...628A..35K}. The early Gaia data release 3 (EDR3) published on 4 December 2020 \citep{2020arXiv201201533G} significantly improved the accuracy of the measured MW Cepheid parallaxes of MW Cepheids. A mitigation of the uncertainty due to the instrumental parallax zero point through an ad hoc position-, color- and magnitude- dependent calibration is also presented by \citet{2020arXiv201201742L}. As discussed by \citet{Riess_2021} (see also Breuval et al.~2021, in prep.), this improvement brings the calibration of Cepheids luminosities to a 1\% level, which makes them the most accurate distance indicators available to date. As the number of measurement epochs and the understanding of the Gaia instrument increase, the DR3 and DR4 will eventually provide trigonometric reliable parallax measurements at a few percent level or better for hundreds of Milky Way Cepheids. Combined with accurate photometry and extinction corrections from 3D extinction maps \citep[see, e.g.,][]{2019MNRAS.483.4277C,2020A&A...641A..79H}, this set of absolutely calibrated distances will result in a very tight set of Cepheid PL relations, calibrated for the solar metallicity.
Our Galaxy therefore appears as a particularly appealing alternative to the Magellanic Clouds as the primary anchor for extragalactic Cepheid distances, thanks to the similarity of its metallicity with those of distant SNeIa host galaxies.
Relying on Milky Way Cepheids presents the advantage of reducing the possible bias introduced by the metallicity correction. This will effectively bypass the metallicity correction, thus increasing the overall robustness of the SNeIa calibration.

As also noted in \citet{Riess2019Nat}, precise measurement of $H_0$ provides a powerful end-to-end test of the LCDM standard model of cosmology. Future observational progress and inspections such as the current study would eventually conclude whether the $H_0$ tension is caused by a measurement error, or it means that the LCDM should be abandoned as a correct model of the Universe.

\acknowledgments
We thank the anonymous referee for the constructive comments. The research leading to these results has received funding from the European Research Council (ERC) under the European Union's Horizon 2020 research and innovation program under grant agreement No 695099 (project CepBin). WG and GP gratefully acknowledge financial support for this work from the BASAL Centro de Astrofisica y Tecnologias Afines (CATA) AFB-170002. GP also acknowledges the support from NCN MAESTRO grant UMO-2017/26/A/ST9/00446 and DIR/WK/2018/09 grants of the Polish Ministry of Science and Higher Education. PK, NN, VH and SB acknowledge the support of the French Agence Nationale de la Recherche (ANR), under grant ANR-15-CE31-0012-01 (project UnlockCepheids). We acknowledge the use of the HST observations of NGC~5584 performed by the SH0ES team (PI: Adam Riess, Cycle: 17, Proposal ID: 11570) which are publicly available on the MAST database. This research made use of Photutils, an Astropy package for detection and photometry of astronomical sources \citep{Bradley2019}.

%


\facility{HST, MAST}
\software{\texttt{DrizzlePac}, Astropy \citep{Bradley2019}, \texttt{scikit-learn}.}



\bibliography{references}{}

\appendix

\section{Aperture Correction}\label{sec:appendix_appcor}
Here, we compare aperture correction using the aperture photometry of the PSF model with the approach using aperture and PSF photometry of stars in the image. For this purpose we compare the PSF and aperture photometry of thirteen uncrowded stars using the stacked version of all the F555W band images. For these stars we crop a 50 pixel $\times$ 50 pixel portion of the image and perform the PSF photometry in the same way as described in Section \ref{sec:psf_phot}. All the sources except from the central star are removed using the PSF modelling prior to the aperture photometry with an aperture radius of 10 pixels. The difference is then measured using Equation \ref{eq:ap_cor} and we find a mean value of $AP_{cor}=0.056\pm0.024$ mag for the F555W band. It is $1\sigma$ larger than the value obtained using aperture photometry of the PSF model. We expect a similar result for F814W band. For the F160W band, the $AP_{cor}$ we measure using the PSF model, i.e. 0.049 mag, is also around $1\sigma$ smaller than the $0.06\pm0.01$ measured by \citet{Huang2020} for F160W images of the SNIa host NGC 1559. The difference of these two methods is most probably due to imperfect subtraction of the background noise in the actual images and the absence of this noise in the PSF model. Therefore, by noting that the difference in the F160W band is most relevant for distance measurement (Equation \ref{eq:wesenheit_mag}), measuring $AP_{cor}$ using aperture photometry of the PSF model rather than using uncrowded stars in the image leads to around $0.01\pm0.01$ mag decrease in the distance modulus.

\begin{figure*}
    \centering
    \includegraphics[width=\textwidth]{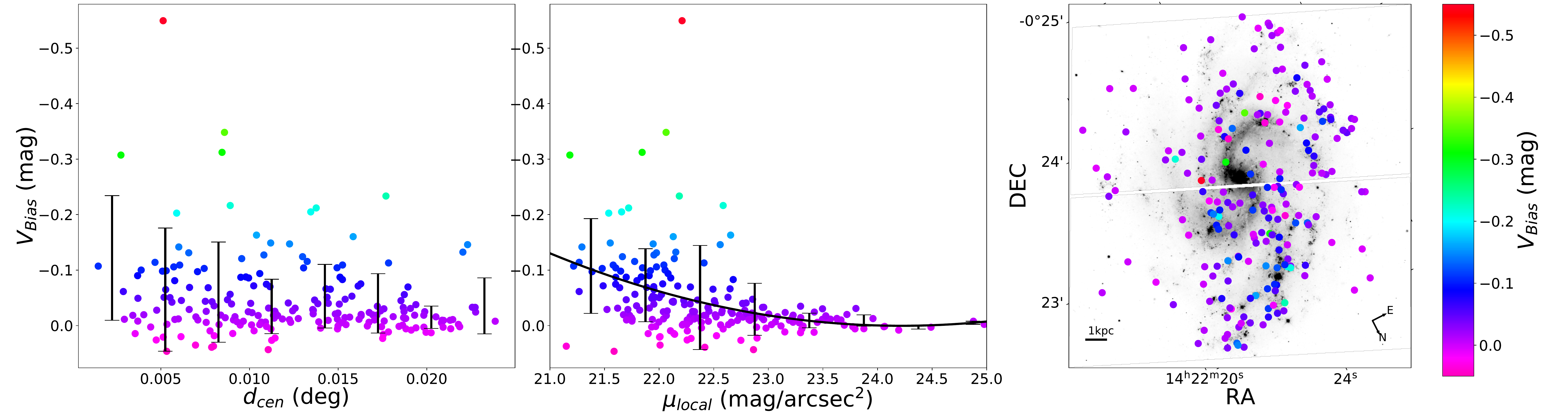}
    \caption{Left: V band crowding bias ($V_{Bias}$) vs. the projected angular distance $d_{cen}$ in degrees from the center of the galaxy. Middle: $V_{Bias}$ vs. local surface brightness $\mu_{local}$. Right: Distribution of the Cepheids in the NGC~5584 with the color code being the estimated crowding bias at the position of each Cepheid in the V band. The same color code is also used in the other panels. The Cepheids that are most affected by the crowding bias are statistically closer to the centre of the galaxy, however, the bias is found to be more correlated with local surface brightness. The linear correlation coefficient between the absolute value of the crowding bias and $d_{cen}$ and $\mu_{local}$ are $r=-0.24$ and $r=-0.40$, respectively. See appendix \ref{sec:appendix_crd} for details.}
    \label{fig:crd_on_gal}
\end{figure*}

\begin{figure}
    \centering
    \includegraphics[scale=0.68]{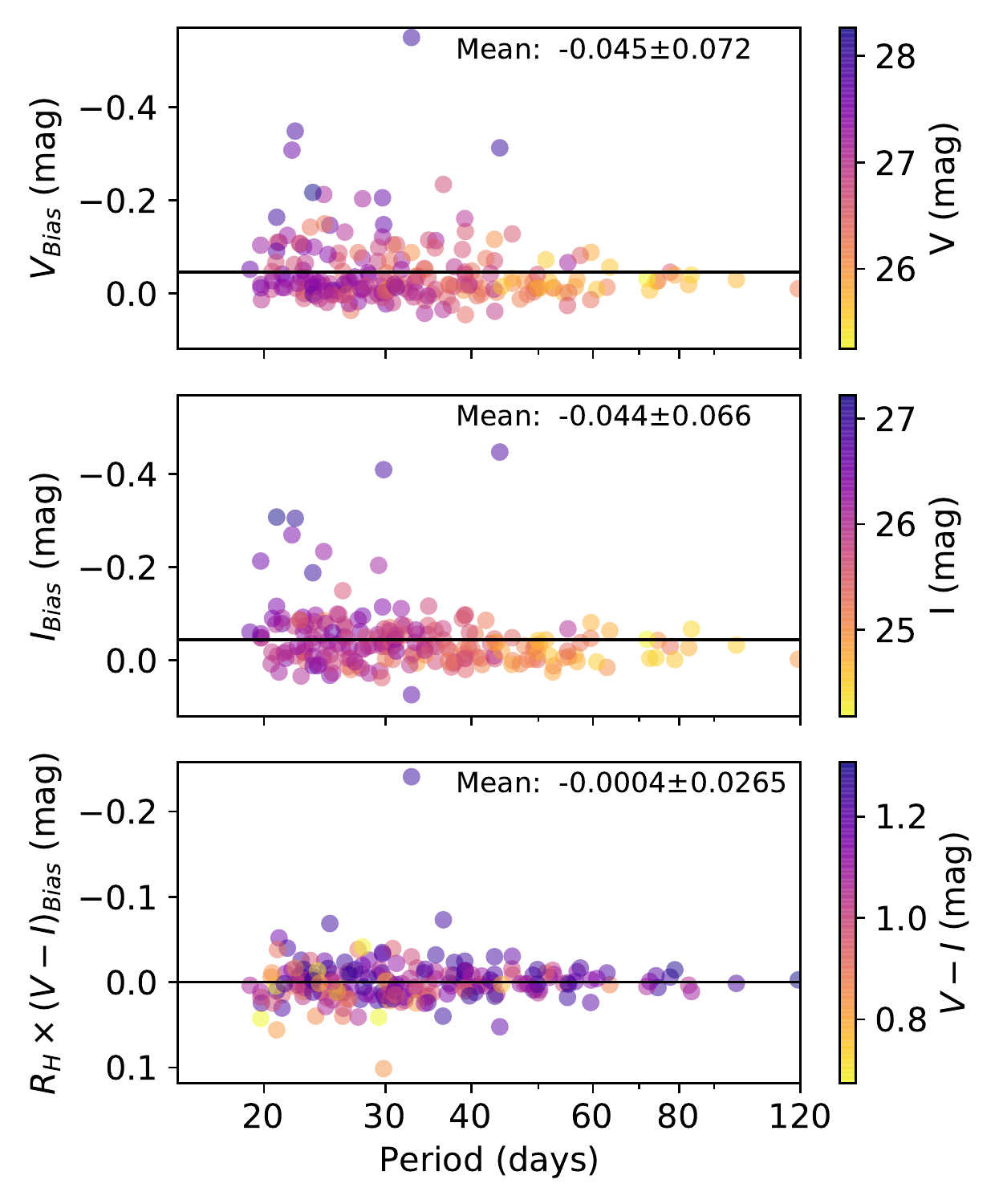}
    \caption{Crowding bias in V and I bands, as well as $R_H(V-I)$ vs. period. This figure can be directly compared to Figure 14 of \citetalias{Hoffmann2016} (see appendix \ref{sec:appendix_crd} for a discussion).}
    \label{fig:crowding_vs_per}
\end{figure}

\section{Crowding Bias}\label{sec:appendix_crd}
In Section \ref{sec:crowding} we explained our method of estimating the crowding bias at the location of each Cepheid. Here we investigate the environmental dependence of crowding bias in the F555W band. Figure \ref{fig:crd_on_gal} shows the distribution of crowding bias across the galaxy (right panel), bias vs. the projected angular distance $d_{cen}$ in degrees from the center of the galaxy (left panel), and bias vs. local surface brightness, $\mu_{local}$ (middle panel). The Cepheids that are most affected by the crowding bias are statistically closer to the centre of the galaxy where the stellar density is generally higher. However, small bias values can also be found at smaller $d_{cen}$ and we measure a correlation coefficient of $r=-0.24$ between the absolute value of the crowding bias and $d_{cen}$. The small correlation is possibly due to spiral structure of NGC 5584, i.e. even at small galactocentric distances, there are less crowded regions.

We also check the crowding bias vs. $\mu_{local}$. The latter is measured using the following three step method: i) we first measure the average sky background using around twenty 40$\times$40 pixel square regions outside the parts of the image covered by NGC 5584, ii) we then used the same size squares at the location of each Cepheid (where we already estimated crowding bias) and measured the total flux inside them, iii) in the end, the average sky background is subtracted from the local total fluxes and the result is converted to surface brightness using the angular area in arcsec and the magnitude zero point. We measure a correlation coefficient of $r=-0.40$ between the crowding bias and $\mu_{local}$. This relatively larger correlation implies that the local surface brightness is a better proxy to crowding bias compared to galactocentric distance. Using a second order polynomial fit we find

\begin{equation}
    \mathrm{Bias}=-0.013\mu_{local}^{2}+0.61\mu_{local}-7.4
\end{equation}
with the standard deviation of the model minus data being $\sigma_{fit}= 0.066$. This relation may be used to estimate the crowding bias in the F555W band from the local surface brightness instead of the artificial star injection approach. It may be useful to note that in the regions with $\mu_{local} \gtrsim 23$ mag/arcsec$^2$, the effect of crowding bias is negligible.

To compare our crowding bias measurements with those of \citetalias{Hoffmann2016}, we also show the bias as a function of period in Figure \ref{fig:crowding_vs_per}. This figure can be directly compared to Figure 14 of \citetalias{Hoffmann2016}. Our results is (within $1\sigma$) in agreement with those of the \citetalias{Hoffmann2016}. In particular, as can be seen in the lowest panel, the effect of the crowding bias on the V-I color measurement is very small. We remind the reader that unlike the approach of SH0ES (averaging over the galaxy for the optical observations), we apply the crowding bias estimated for individual Cepheids on their photometric results before the template fitting.

\newpage

\section{Cepheid Properties}
Our results for the photometric properties of the identified Cepheids in NGC~5584 are listed in Table \ref{tab:all_ceph_results}.

\startlongtable
\begin{deluxetable*}{cccccccccccccc}
\tabletypesize{\scriptsize}



\tablecaption{Our measurements for the photometric properties of the Cepheids in the NGC~5584. \label{tab:all_ceph_results}}


\tablehead{\colhead{ID} & \colhead{RAJ2000} & \colhead{DECJ2000} & \colhead{LP} & \colhead{$\sigma_{LP}$} & \colhead{$A_{LP}$} & \colhead{$V$} & \colhead{$\sigma_V$} & \colhead{$A_V$} & \colhead{$I$} & \colhead{$\sigma_I$} & \colhead{$A_I$} & \colhead{$H$} & \colhead{$\sigma_{H}$} \\ 
\colhead{} & \colhead{(deg)} & \colhead{(deg)} & \colhead{(mag)} & \colhead{(mag)} & \colhead{(mag)} & \colhead{(mag)} & \colhead{(mag)} & \colhead{(mag)} & \colhead{(mag)} & \colhead{(mag)} & \colhead{(mag)} & \colhead{(mag)} & \colhead{(mag)} } 

\startdata
82928 & 215.5872 & -0.3685 & 26.717 & 0.019 & 0.313 & 26.943 & 0.024 & 0.36 & 25.899 & 0.044 & 0.231 & 24.859 & 0.1 \\
86318 & 215.5892 & -0.3676 & 26.985 & 0.023 & 0.274 & 27.286 & 0.025 & 0.326 & 26.04 & 0.024 & 0.22 & - & - \\
91999 & 215.5886 & -0.3682 & 27.392 & 0.047 & 0.306 & 27.696 & 0.046 & 0.361 & 26.442 & 0.05 & 0.243 & - & - \\
96368 & 215.584 & -0.3708 & 26.841 & 0.018 & 0.214 & 27.11 & 0.023 & 0.255 & 25.945 & 0.039 & 0.17 & - & - \\
97566 & 215.5886 & -0.3685 & 26.3 & 0.019 & 0.203 & 26.6 & 0.02 & 0.244 & 25.35 & 0.019 & 0.166 & 24.107 & 0.023 \\
111577 & 215.5875 & -0.3698 & 25.58 & 0.008 & 0.18 & 25.793 & 0.012 & 0.21 & 24.773 & 0.017 & 0.133 & 23.775 & 0.042 \\
121760 & 215.588 & -0.3701 & 27.399 & 0.059 & 0.29 & 27.701 & 0.07 & 0.343 & 26.453 & 0.107 & 0.231 & - & - \\
134935 & 215.5855 & -0.372 & 25.963 & 0.011 & 0.232 & 26.213 & 0.015 & 0.272 & 25.099 & 0.02 & 0.179 & - & - \\
143986 & 215.5898 & -0.3703 & 25.809 & 0.019 & 0.168 & 25.984 & 0.028 & 0.192 & 25.067 & 0.026 & 0.116 & 24.187 & 0.065 \\
151156 & 215.5928 & -0.3691 & 27.043 & 0.055 & 0.186 & 27.282 & 0.074 & 0.22 & 26.193 & 0.07 & 0.143 & - & - \\
156158 & 215.5903 & -0.3706 & 26.541 & 0.019 & 0.245 & 26.715 & 0.026 & 0.277 & 25.808 & 0.031 & 0.171 & - & - \\
157119 & 215.5914 & -0.3701 & 25.616 & 0.016 & 0.234 & 25.824 & 0.019 & 0.269 & 24.823 & 0.027 & 0.171 & 23.838 & 0.059 \\
172880 & 215.589 & -0.3721 & 25.385 & 0.036 & 0.172 & 25.673 & 0.04 & 0.205 & 24.444 & 0.039 & 0.128 & 23.24 & 0.062 \\
175404 & 215.591 & -0.3712 & 26.232 & 0.018 & 0.182 & 26.528 & 0.019 & 0.221 & 25.284 & 0.016 & 0.149 & - & - \\
175413 & 215.5939 & -0.3697 & 26.399 & 0.02 & 0.206 & 26.69 & 0.026 & 0.25 & 25.463 & 0.026 & 0.169 & 24.233 & 0.061 \\
185292 & 215.5952 & -0.3696 & 25.475 & 0.008 & 0.211 & 25.635 & 0.012 & 0.237 & 24.761 & 0.009 & 0.144 & 23.914 & 0.025 \\
191706 & 215.5961 & -0.3695 & 26.359 & 0.017 & 0.17 & 26.569 & 0.023 & 0.198 & 25.558 & 0.029 & 0.124 & 24.575 & 0.067 \\
197260 & 215.59 & -0.3729 & 27.219 & 0.03 & 0.26 & 27.521 & 0.03 & 0.308 & 26.268 & 0.031 & 0.208 & - & - \\
200467 & 215.5944 & -0.3707 & 26.947 & 0.033 & 0.285 & 27.253 & 0.033 & 0.338 & 25.992 & 0.034 & 0.227 & 24.72 & 0.045 \\
200686 & 215.5899 & -0.3731 & 25.929 & 0.022 & 0.186 & 26.234 & 0.024 & 0.229 & 24.97 & 0.024 & 0.157 & - & - \\
208725 & 215.5952 & -0.3708 & 26.322 & 0.016 & 0.266 & 26.524 & 0.021 & 0.304 & 25.543 & 0.027 & 0.193 & - & - \\
211148 & 215.5834 & -0.3769 & 27.295 & 0.035 & 0.299 & 27.546 & 0.043 & 0.347 & 26.435 & 0.076 & 0.226 & - & - \\
216328 & 215.5967 & -0.3704 & 26.468 & 0.037 & 0.21 & 26.767 & 0.038 & 0.255 & 25.52 & 0.038 & 0.176 & - & - \\
220248 & 215.5879 & -0.3751 & 26.293 & 0.016 & 0.294 & 26.526 & 0.021 & 0.339 & 25.462 & 0.023 & 0.219 & 24.399 & 0.051 \\
229600 & 215.5929 & -0.373 & 26.922 & 0.027 & 0.252 & 27.162 & 0.032 & 0.293 & 26.076 & 0.041 & 0.191 & - & - \\
230093 & 215.5895 & -0.3747 & 26.99 & 0.017 & 0.207 & 27.263 & 0.021 & 0.248 & 26.085 & 0.029 & 0.165 & 24.914 & 0.067 \\
238461 & 215.5937 & -0.373 & 26.733 & 0.022 & 0.268 & 27.037 & 0.023 & 0.319 & 25.781 & 0.021 & 0.216 & 24.51 & 0.023 \\
247527 & 215.5757 & -0.3826 & 26.551 & 0.019 & 0.296 & 26.801 & 0.024 & 0.343 & 25.691 & 0.037 & 0.225 & - & - \\
253461 & 215.5963 & -0.3724 & 26.342 & 0.018 & 0.193 & 26.636 & 0.018 & 0.235 & 25.4 & 0.019 & 0.16 & 24.163 & 0.024 \\
254240 & 215.6057 & -0.3677 & 26.154 & 0.013 & 0.319 & 26.418 & 0.016 & 0.371 & 25.27 & 0.023 & 0.243 & - & - \\
258671 & 215.5955 & -0.3731 & 27.075 & 0.04 & 0.223 & 27.349 & 0.051 & 0.266 & 26.172 & 0.081 & 0.179 & - & - \\
267902 & 215.5968 & -0.373 & 25.861 & 0.014 & 0.227 & 26.066 & 0.02 & 0.261 & 25.072 & 0.021 & 0.166 & - & - \\
271193 & 215.5966 & -0.3732 & 26.321 & 0.019 & 0.171 & 26.569 & 0.028 & 0.204 & 25.456 & 0.029 & 0.133 & - & - \\
271677 & 215.5882 & -0.3775 & 27.113 & 0.024 & 0.23 & 27.413 & 0.024 & 0.277 & 26.166 & 0.025 & 0.188 & - & - \\
276835 & 215.5827 & -0.3806 & 26.582 & 0.036 & 0.243 & 26.696 & 0.047 & 0.266 & 25.948 & 0.068 & 0.157 & - & - \\
281768 & 215.5913 & -0.3764 & 27.006 & 0.02 & 0.284 & 27.277 & 0.028 & 0.331 & 26.111 & 0.039 & 0.219 & 24.945 & 0.092 \\
290494 & 215.6083 & -0.3682 & 26.103 & 0.012 & 0.267 & 26.304 & 0.015 & 0.305 & 25.322 & 0.025 & 0.195 & - & - \\
295981 & 215.5898 & -0.3779 & 25.913 & 0.013 & 0.193 & 26.104 & 0.016 & 0.221 & 25.145 & 0.027 & 0.138 & 24.21 & 0.062 \\
298430 & 215.6007 & -0.3724 & 26.845 & 0.024 & 0.41 & 27.071 & 0.03 & 0.469 & 26.031 & 0.047 & 0.302 & 24.993 & 0.107 \\
321323 & 215.5979 & -0.375 & 26.971 & 0.038 & 0.218 & 27.269 & 0.037 & 0.262 & 26.027 & 0.041 & 0.178 & - & - \\
321793 & 215.5995 & -0.3742 & 26.995 & 0.031 & 0.196 & 27.164 & 0.038 & 0.221 & 26.266 & 0.062 & 0.135 & - & - \\
325206 & 215.5894 & -0.3795 & 25.933 & 0.018 & 0.207 & 26.226 & 0.024 & 0.25 & 24.994 & 0.029 & 0.17 & 23.76 & 0.069 \\
325458 & 215.5992 & -0.3745 & 27.414 & 0.102 & 0.286 & 27.723 & 0.103 & 0.338 & 26.453 & 0.102 & 0.228 & - & - \\
325693 & 215.5909 & -0.3788 & 27.443 & 0.056 & 0.264 & 27.636 & 0.074 & 0.3 & 26.678 & 0.113 & 0.191 & - & - \\
325718 & 215.5996 & -0.3744 & 25.435 & 0.011 & 0.178 & 25.726 & 0.011 & 0.215 & 24.492 & 0.011 & 0.14 & 23.272 & 0.013 \\
326705 & 215.5967 & -0.3759 & 27.603 & 0.037 & 0.443 & 27.754 & 0.046 & 0.5 & 26.917 & 0.075 & 0.312 & - & - \\
329366 & 215.6014 & -0.3736 & 26.959 & 0.028 & 0.193 & 27.255 & 0.029 & 0.235 & 26.012 & 0.03 & 0.159 & 24.771 & 0.039 \\
330805 & 215.599 & -0.3749 & 26.173 & 0.015 & 0.183 & 26.361 & 0.02 & 0.21 & 25.411 & 0.021 & 0.13 & - & - \\
339133 & 215.5806 & -0.3847 & 25.609 & 0.009 & 0.185 & 25.813 & 0.013 & 0.213 & 24.817 & 0.014 & 0.135 & - & - \\
340379 & 215.5949 & -0.3775 & 26.966 & 0.055 & 0.203 & 27.119 & 0.076 & 0.227 & 26.263 & 0.1 & 0.137 & 25.44 & 0.237 \\
342112 & 215.5925 & -0.3788 & 26.002 & 0.015 & 0.175 & 26.278 & 0.019 & 0.212 & 25.09 & 0.025 & 0.141 & - & - \\
347072 & 215.5997 & -0.3754 & 25.593 & 0.018 & 0.164 & 25.803 & 0.02 & 0.191 & 24.791 & 0.044 & 0.12 & - & - \\
353561 & 215.594 & -0.3786 & 26.399 & 0.046 & 0.164 & 26.642 & 0.064 & 0.195 & 25.54 & 0.096 & 0.126 & - & - \\
354807 & 215.594 & -0.3787 & 25.535 & 0.018 & 0.159 & 25.757 & 0.026 & 0.184 & 24.707 & 0.026 & 0.111 & 23.699 & 0.064 \\
374736 & 215.5992 & -0.377 & 26.048 & 0.013 & 0.242 & 26.286 & 0.015 & 0.281 & 25.205 & 0.024 & 0.183 & 24.136 & 0.055 \\
378235 & 215.6091 & -0.3721 & 26.737 & 0.016 & 0.264 & 26.989 & 0.022 & 0.308 & 25.873 & 0.028 & 0.201 & 24.76 & 0.065 \\
390652 & 215.6005 & -0.3771 & 26.415 & 0.021 & 0.239 & 26.603 & 0.026 & 0.272 & 25.656 & 0.037 & 0.17 & 24.731 & 0.084 \\
395114 & 215.5969 & -0.3792 & 25.537 & 0.017 & 0.184 & 25.853 & 0.019 & 0.228 & 24.557 & 0.022 & 0.157 & 23.258 & 0.04 \\
396420 & 215.6059 & -0.3747 & 26.928 & 0.018 & 0.213 & 27.224 & 0.018 & 0.258 & 25.984 & 0.019 & 0.177 & - & - \\
399436 & 215.5982 & -0.3788 & 26.727 & 0.029 & 0.332 & 26.893 & 0.04 & 0.375 & 26.011 & 0.044 & 0.233 & - & - \\
411135 & 215.597 & -0.3799 & 25.921 & 0.012 & 0.281 & 26.153 & 0.017 & 0.325 & 25.091 & 0.025 & 0.21 & 24.039 & 0.062 \\
412396 & 215.6 & -0.3785 & 26.137 & 0.023 & 0.192 & 26.308 & 0.033 & 0.218 & 25.403 & 0.044 & 0.133 & 24.528 & 0.105 \\
418643 & 215.5894 & -0.3842 & 26.447 & 0.016 & 0.3 & 26.743 & 0.023 & 0.353 & 25.511 & 0.03 & 0.236 & - & - \\
419182 & 215.5993 & -0.3792 & 26.311 & 0.023 & 0.196 & 26.476 & 0.027 & 0.221 & 25.587 & 0.048 & 0.134 & - & - \\
420418 & 215.5948 & -0.3815 & 26.579 & 0.026 & 0.219 & 26.83 & 0.035 & 0.258 & 25.712 & 0.037 & 0.17 & - & - \\
421192 & 215.5971 & -0.3804 & 26.244 & 0.009 & 0.362 & 26.511 & 0.012 & 0.419 & 25.36 & 0.017 & 0.273 & - & - \\
424677 & 215.5991 & -0.3795 & 27.785 & 0.033 & 0.158 & 28.065 & 0.045 & 0.192 & 26.863 & 0.07 & 0.125 & - & - \\
427599 & 215.5982 & -0.3801 & 26.055 & 0.021 & 0.178 & 26.269 & 0.028 & 0.208 & 25.25 & 0.033 & 0.131 & - & - \\
437977 & 215.5933 & -0.3831 & 27.197 & 0.053 & 0.234 & 27.283 & 0.074 & 0.253 & 26.609 & 0.082 & 0.145 & - & - \\
446943 & 215.5947 & -0.3829 & 25.601 & 0.011 & 0.212 & 25.858 & 0.014 & 0.252 & 24.725 & 0.019 & 0.166 & - & - \\
449157 & 215.5933 & -0.3837 & 26.347 & 0.018 & 0.169 & 26.578 & 0.025 & 0.199 & 25.511 & 0.026 & 0.127 & 24.467 & 0.061 \\
449432 & 215.6042 & -0.3782 & 25.017 & 0.012 & 0.197 & 25.248 & 0.016 & 0.23 & 24.18 & 0.013 & 0.15 & 23.139 & 0.031 \\
455910 & 215.6028 & -0.3792 & 27.031 & 0.033 & 0.216 & 27.304 & 0.041 & 0.258 & 26.126 & 0.053 & 0.174 & - & - \\
455911 & 215.6042 & -0.3785 & 26.348 & 0.016 & 0.276 & 26.63 & 0.022 & 0.324 & 25.433 & 0.022 & 0.216 & - & - \\
464626 & 215.5896 & -0.3864 & 26.954 & 0.029 & 0.371 & 27.26 & 0.038 & 0.432 & 26.002 & 0.045 & 0.287 & 24.728 & 0.102 \\
466137 & 215.6009 & -0.3807 & 27.027 & 0.023 & 0.33 & 27.334 & 0.025 & 0.388 & 26.072 & 0.023 & 0.26 & - & - \\
469580 & 215.5999 & -0.3814 & 26.119 & 0.015 & 0.177 & 26.358 & 0.022 & 0.21 & 25.269 & 0.025 & 0.136 & - & - \\
473829 & 215.6056 & -0.3787 & 25.629 & 0.012 & 0.201 & 25.906 & 0.015 & 0.243 & 24.716 & 0.02 & 0.164 & 23.527 & 0.045 \\
475792 & 215.5941 & -0.3846 & 27.131 & 0.086 & 0.379 & 27.223 & 0.088 & 0.421 & 26.541 & 0.09 & 0.254 & - & - \\
477073 & 215.6036 & -0.3799 & 26.683 & 0.025 & 0.205 & 26.822 & 0.036 & 0.228 & 26.004 & 0.04 & 0.135 & - & - \\
478350 & 215.602 & -0.3807 & 26.604 & 0.028 & 0.289 & 26.915 & 0.03 & 0.34 & 25.638 & 0.029 & 0.23 & 24.359 & 0.041 \\
481285 & 215.5936 & -0.3852 & 26.15 & 0.017 & 0.202 & 26.342 & 0.025 & 0.232 & 25.383 & 0.02 & 0.145 & - & - \\
487089 & 215.5934 & -0.3855 & 26.549 & 0.029 & 0.217 & 26.72 & 0.038 & 0.246 & 25.816 & 0.053 & 0.151 & - & - \\
491027 & 215.5998 & -0.3825 & 27.04 & 0.037 & 0.389 & 27.245 & 0.046 & 0.443 & 26.261 & 0.088 & 0.282 & - & - \\
493790 & 215.5985 & -0.3833 & 26.514 & 0.016 & 0.203 & 26.72 & 0.025 & 0.235 & 25.722 & 0.029 & 0.149 & 24.748 & 0.071 \\
494049 & 215.6008 & -0.3821 & 26.727 & 0.032 & 0.209 & 26.918 & 0.049 & 0.24 & 25.962 & 0.049 & 0.149 & - & - \\
495038 & 215.5946 & -0.3853 & 26.722 & 0.026 & 0.224 & 26.807 & 0.026 & 0.242 & 26.132 & 0.026 & 0.138 & - & - \\
502797 & 215.6009 & -0.3825 & 25.936 & 0.015 & 0.274 & 26.2 & 0.017 & 0.321 & 25.052 & 0.029 & 0.211 & 23.902 & 0.064 \\
504490 & 215.5963 & -0.3849 & 26.289 & 0.017 & 0.237 & 26.475 & 0.02 & 0.27 & 25.533 & 0.032 & 0.169 & 24.614 & 0.071 \\
511109 & 215.6 & -0.3833 & 26.693 & 0.022 & 0.384 & 26.982 & 0.027 & 0.445 & 25.772 & 0.042 & 0.293 & 24.54 & 0.097 \\
513372 & 215.6028 & -0.382 & 26.463 & 0.013 & 0.342 & 26.73 & 0.019 & 0.396 & 25.576 & 0.023 & 0.26 & 24.417 & 0.056 \\
513827 & 215.5974 & -0.3849 & 26.571 & 0.023 & 0.31 & 26.878 & 0.024 & 0.366 & 25.617 & 0.022 & 0.246 & 24.335 & 0.025 \\
516608 & 215.596 & -0.3857 & 27.256 & 0.031 & 0.221 & 27.478 & 0.048 & 0.257 & 26.44 & 0.067 & 0.165 & - & - \\
519642 & 215.5948 & -0.3864 & 26.551 & 0.014 & 0.335 & 26.761 & 0.016 & 0.383 & 25.759 & 0.026 & 0.245 & - & - \\
521128 & 215.5939 & -0.387 & 26.545 & 0.024 & 0.235 & 26.832 & 0.028 & 0.28 & 25.62 & 0.044 & 0.188 & 24.405 & 0.096 \\
534937 & 215.5823 & -0.3936 & 27.038 & 0.039 & 0.306 & 27.277 & 0.047 & 0.354 & 26.199 & 0.076 & 0.228 & 25.115 & 0.169 \\
540558 & 215.5994 & -0.3851 & 26.528 & 0.036 & 0.201 & 26.689 & 0.046 & 0.226 & 25.811 & 0.064 & 0.138 & - & - \\
543151 & 215.6031 & -0.3834 & 26.431 & 0.021 & 0.208 & 26.732 & 0.022 & 0.25 & 25.479 & 0.022 & 0.17 & 24.231 & 0.029 \\
549082 & 215.5961 & -0.3872 & 25.888 & 0.011 & 0.207 & 26.132 & 0.016 & 0.244 & 25.035 & 0.018 & 0.159 & 23.947 & 0.044 \\
549585 & 215.5937 & -0.3885 & 26.751 & 0.019 & 0.345 & 26.926 & 0.03 & 0.39 & 26.021 & 0.041 & 0.244 & 25.13 & 0.101 \\
550433 & 215.6125 & -0.3789 & 26.398 & 0.015 & 0.342 & 26.691 & 0.022 & 0.399 & 25.468 & 0.027 & 0.265 & 24.227 & 0.067 \\
550434 & 215.6129 & -0.3787 & 27.084 & 0.022 & 0.349 & 27.391 & 0.022 & 0.408 & 26.132 & 0.023 & 0.272 & 24.848 & 0.03 \\
552392 & 215.5834 & -0.3939 & 26.655 & 0.061 & 0.188 & 26.872 & 0.064 & 0.219 & 25.843 & 0.121 & 0.139 & - & - \\
556696 & 215.6031 & -0.384 & 26.784 & 0.022 & 0.214 & 27.081 & 0.023 & 0.258 & 25.839 & 0.023 & 0.175 & - & - \\
562692 & 215.6024 & -0.3847 & 26.409 & 0.03 & 0.384 & 26.659 & 0.033 & 0.441 & 25.555 & 0.058 & 0.285 & - & - \\
562960 & 215.6016 & -0.3851 & 27.236 & 0.041 & 0.389 & 27.548 & 0.04 & 0.452 & 26.275 & 0.045 & 0.301 & - & - \\
563696 & 215.6054 & -0.3832 & 27.128 & 0.041 & 0.354 & 27.361 & 0.048 & 0.406 & 26.301 & 0.057 & 0.262 & 25.241 & 0.118 \\
567349 & 215.6029 & -0.3846 & 25.978 & 0.02 & 0.191 & 26.264 & 0.025 & 0.232 & 25.05 & 0.039 & 0.157 & - & - \\
571414 & 215.6052 & -0.3837 & 26.151 & 0.016 & 0.314 & 26.445 & 0.021 & 0.367 & 25.218 & 0.032 & 0.245 & 23.982 & 0.077 \\
584459 & 215.5942 & -0.3899 & 27.753 & 0.094 & 0.289 & 28.06 & 0.101 & 0.342 & 26.795 & 0.089 & 0.23 & - & - \\
584466 & 215.5955 & -0.3893 & 26.698 & 0.036 & 0.208 & 26.844 & 0.048 & 0.232 & 26.008 & 0.068 & 0.139 & - & - \\
589456 & 215.6036 & -0.3854 & 26.996 & 0.039 & 0.188 & 27.29 & 0.04 & 0.229 & 26.055 & 0.038 & 0.155 & 24.823 & 0.037 \\
594530 & 215.5833 & -0.396 & 26.777 & 0.017 & 0.191 & 27.021 & 0.021 & 0.225 & 25.922 & 0.04 & 0.147 & - & - \\
602554 & 215.6093 & -0.3831 & 26.817 & 0.019 & 0.365 & 27.126 & 0.02 & 0.427 & 25.863 & 0.02 & 0.284 & 24.57 & 0.026 \\
603762 & 215.6004 & -0.3877 & 26.322 & 0.022 & 0.176 & 26.618 & 0.021 & 0.216 & 25.374 & 0.022 & 0.146 & - & - \\
605531 & 215.593 & -0.3916 & 26.61 & 0.017 & 0.257 & 26.858 & 0.021 & 0.3 & 25.753 & 0.042 & 0.196 & 24.646 & 0.096 \\
606041 & 215.6089 & -0.3835 & 25.872 & 0.012 & 0.184 & 26.059 & 0.016 & 0.21 & 25.11 & 0.018 & 0.13 & 24.19 & 0.042 \\
607520 & 215.6078 & -0.3841 & 25.387 & 0.012 & 0.226 & 25.648 & 0.015 & 0.267 & 24.503 & 0.02 & 0.177 & 23.368 & 0.047 \\
610213 & 215.604 & -0.3862 & 26.119 & 0.011 & 0.198 & 26.413 & 0.016 & 0.24 & 25.18 & 0.018 & 0.164 & 23.944 & 0.043 \\
611528 & 215.6102 & -0.3831 & 26.571 & 0.03 & 0.239 & 26.763 & 0.039 & 0.273 & 25.805 & 0.058 & 0.171 & - & - \\
620130 & 215.5825 & -0.3977 & 26.604 & 0.016 & 0.3 & 26.868 & 0.02 & 0.349 & 25.723 & 0.023 & 0.229 & - & - \\
628911 & 215.6084 & -0.3849 & 25.437 & 0.012 & 0.178 & 25.685 & 0.015 & 0.212 & 24.573 & 0.023 & 0.138 & 23.476 & 0.052 \\
633407 & 215.6025 & -0.3881 & 26.726 & 0.02 & 0.176 & 27.016 & 0.02 & 0.215 & 25.788 & 0.021 & 0.145 & - & - \\
640109 & 215.5982 & -0.3906 & 27.413 & 0.03 & 0.18 & 27.663 & 0.042 & 0.214 & 26.547 & 0.069 & 0.14 & - & - \\
644384 & 215.5912 & -0.3945 & 26.944 & 0.024 & 0.357 & 27.171 & 0.03 & 0.409 & 26.126 & 0.042 & 0.263 & - & - \\
648122 & 215.5924 & -0.394 & 27.972 & 0.045 & 0.173 & 28.263 & 0.045 & 0.213 & 27.032 & 0.045 & 0.143 & - & - \\
648136 & 215.5955 & -0.3924 & 26.563 & 0.046 & 0.158 & 26.855 & 0.047 & 0.193 & 25.621 & 0.045 & 0.127 & - & - \\
656817 & 215.609 & -0.386 & 25.391 & 0.012 & 0.169 & 25.566 & 0.016 & 0.193 & 24.648 & 0.021 & 0.117 & 23.767 & 0.05 \\
668576 & 215.5936 & -0.3944 & 27.419 & 0.045 & 0.186 & 27.711 & 0.046 & 0.227 & 26.479 & 0.045 & 0.154 & - & - \\
673309 & 215.5921 & -0.3954 & 26.015 & 0.018 & 0.186 & 26.313 & 0.018 & 0.227 & 25.067 & 0.018 & 0.155 & 23.825 & 0.021 \\
673828 & 215.6087 & -0.387 & 25.579 & 0.013 & 0.168 & 25.833 & 0.018 & 0.201 & 24.702 & 0.015 & 0.131 & - & - \\
674808 & 215.6013 & -0.3908 & 27.467 & 0.032 & 0.262 & 27.768 & 0.032 & 0.311 & 26.519 & 0.033 & 0.21 & - & - \\
696165 & 215.6089 & -0.388 & 26.487 & 0.016 & 0.252 & 26.64 & 0.019 & 0.283 & 25.787 & 0.029 & 0.172 & - & - \\
697115 & 215.6045 & -0.3902 & 27.061 & 0.028 & 0.193 & 27.351 & 0.034 & 0.235 & 26.127 & 0.05 & 0.159 & - & - \\
708572 & 215.5999 & -0.3932 & 26.214 & 0.018 & 0.33 & 26.408 & 0.025 & 0.375 & 25.45 & 0.034 & 0.236 & - & - \\
711358 & 215.6128 & -0.3867 & 26.85 & 0.021 & 0.327 & 27.106 & 0.029 & 0.379 & 25.981 & 0.044 & 0.248 & - & - \\
715226 & 215.5996 & -0.3937 & 26.214 & 0.017 & 0.194 & 26.509 & 0.017 & 0.236 & 25.269 & 0.017 & 0.161 & 24.028 & 0.02 \\
715986 & 215.6103 & -0.3882 & 26.597 & 0.023 & 0.298 & 26.906 & 0.024 & 0.351 & 25.637 & 0.025 & 0.236 & 24.36 & 0.037 \\
718451 & 215.5968 & -0.3952 & 26.225 & 0.082 & 0.278 & 26.43 & 0.107 & 0.319 & 25.439 & 0.12 & 0.202 & - & - \\
727892 & 215.6143 & -0.3868 & 26.128 & 0.013 & 0.305 & 26.362 & 0.019 & 0.352 & 25.298 & 0.019 & 0.227 & 24.233 & 0.048 \\
729270 & 215.5882 & -0.4002 & 26.763 & 0.013 & 0.36 & 27.005 & 0.016 & 0.414 & 25.923 & 0.028 & 0.267 & - & - \\
735368 & 215.614 & -0.3873 & 25.286 & 0.016 & 0.189 & 25.53 & 0.025 & 0.223 & 24.429 & 0.03 & 0.146 & 23.349 & 0.076 \\
735776 & 215.5832 & -0.403 & 26.776 & 0.019 & 0.302 & 27.007 & 0.024 & 0.348 & 25.949 & 0.03 & 0.225 & - & - \\
738261 & 215.5993 & -0.3949 & 26.753 & 0.021 & 0.339 & 27.063 & 0.021 & 0.397 & 25.795 & 0.022 & 0.265 & 24.509 & 0.031 \\
740028 & 215.601 & -0.3942 & 26.777 & 0.03 & 0.238 & 27.077 & 0.03 & 0.286 & 25.831 & 0.03 & 0.194 & 24.577 & 0.038 \\
741044 & 215.6151 & -0.387 & 27.273 & 0.036 & 0.233 & 27.48 & 0.048 & 0.268 & 26.482 & 0.078 & 0.17 & - & - \\
744518 & 215.6091 & -0.3902 & 27.905 & 0.098 & 0.351 & 28.051 & 0.118 & 0.392 & 27.225 & 0.127 & 0.24 & - & - \\
757081 & 215.6052 & -0.3929 & 26.076 & 0.015 & 0.242 & 26.334 & 0.019 & 0.284 & 25.2 & 0.024 & 0.187 & - & - \\
758598 & 215.607 & -0.3921 & 26.075 & 0.02 & 0.247 & 26.365 & 0.027 & 0.294 & 25.145 & 0.029 & 0.197 & - & - \\
762356 & 215.6121 & -0.3896 & 25.773 & 0.012 & 0.201 & 26.071 & 0.012 & 0.243 & 24.825 & 0.012 & 0.165 & - & - \\
763038 & 215.5999 & -0.3959 & 26.461 & 0.021 & 0.273 & 26.65 & 0.029 & 0.312 & 25.7 & 0.044 & 0.197 & 24.77 & 0.103 \\
766511 & 215.616 & -0.3879 & 26.237 & 0.011 & 0.258 & 26.498 & 0.015 & 0.303 & 25.358 & 0.02 & 0.199 & 24.219 & 0.047 \\
767732 & 215.6132 & -0.3893 & 26.605 & 0.025 & 0.231 & 26.696 & 0.028 & 0.25 & 26.007 & 0.047 & 0.144 & - & - \\
770504 & 215.6101 & -0.3911 & 26.921 & 0.039 & 0.175 & 27.211 & 0.05 & 0.214 & 25.984 & 0.058 & 0.144 & - & - \\
770520 & 215.6157 & -0.3882 & 25.923 & 0.011 & 0.166 & 26.215 & 0.011 & 0.203 & 24.982 & 0.011 & 0.135 & - & - \\
775000 & 215.6133 & -0.3896 & 26.341 & 0.015 & 0.216 & 26.641 & 0.016 & 0.262 & 25.392 & 0.015 & 0.179 & 24.135 & 0.02 \\
781327 & 215.606 & -0.3937 & 25.326 & 0.022 & 0.156 & 25.545 & 0.03 & 0.183 & 24.506 & 0.029 & 0.114 & - & - \\
781586 & 215.6048 & -0.3943 & 25.501 & 0.016 & 0.179 & 25.625 & 0.02 & 0.197 & 24.845 & 0.025 & 0.114 & - & - \\
784855 & 215.6052 & -0.3943 & 26.72 & 0.023 & 0.244 & 27.021 & 0.023 & 0.293 & 25.773 & 0.022 & 0.199 & - & - \\
787283 & 215.6034 & -0.3953 & 27.786 & 0.116 & 0.43 & 27.946 & 0.171 & 0.485 & 27.085 & 0.101 & 0.303 & 26.234 & 0.268 \\
789264 & 215.6083 & -0.3929 & 26.913 & 0.054 & 0.229 & 27.053 & 0.067 & 0.255 & 26.235 & 0.107 & 0.154 & - & - \\
789518 & 215.6123 & -0.3909 & 26.569 & 0.028 & 0.327 & 26.719 & 0.023 & 0.367 & 25.879 & 0.06 & 0.226 & - & - \\
797934 & 215.613 & -0.391 & 27.463 & 0.048 & 0.331 & 27.702 & 0.063 & 0.381 & 26.625 & 0.077 & 0.246 & - & - \\
801059 & 215.5983 & -0.3986 & 25.378 & 0.013 & 0.188 & 25.64 & 0.018 & 0.225 & 24.49 & 0.024 & 0.15 & 23.35 & 0.057 \\
810216 & 215.6005 & -0.3979 & 26.559 & 0.025 & 0.187 & 26.805 & 0.033 & 0.221 & 25.699 & 0.029 & 0.144 & 24.608 & 0.065 \\
810479 & 215.6089 & -0.3936 & 27.076 & 0.024 & 0.383 & 27.272 & 0.03 & 0.435 & 26.314 & 0.049 & 0.275 & - & - \\
811974 & 215.614 & -0.3911 & 26.005 & 0.011 & 0.257 & 26.213 & 0.019 & 0.295 & 25.214 & 0.019 & 0.189 & 24.22 & 0.051 \\
823580 & 215.6078 & -0.3949 & 26.309 & 0.016 & 0.212 & 26.612 & 0.016 & 0.258 & 25.355 & 0.018 & 0.177 & 24.09 & 0.027 \\
825506 & 215.5976 & -0.4002 & 27.509 & 0.031 & 0.326 & 27.726 & 0.035 & 0.373 & 26.706 & 0.05 & 0.238 & - & - \\
835494 & 215.6062 & -0.3963 & 26.917 & 0.021 & 0.318 & 27.167 & 0.03 & 0.368 & 26.059 & 0.038 & 0.24 & - & - \\
835998 & 215.6124 & -0.3932 & 25.91 & 0.03 & 0.229 & 26.231 & 0.03 & 0.276 & 24.924 & 0.035 & 0.19 & 23.611 & 0.051 \\
845553 & 215.6017 & -0.3991 & 26.156 & 0.017 & 0.296 & 26.371 & 0.021 & 0.339 & 25.357 & 0.033 & 0.216 & - & - \\
845788 & 215.604 & -0.3979 & 26.537 & 0.023 & 0.214 & 26.646 & 0.025 & 0.234 & 25.908 & 0.051 & 0.136 & - & - \\
852752 & 215.6107 & -0.3948 & 26.762 & 0.017 & 0.393 & 27.064 & 0.022 & 0.457 & 25.821 & 0.026 & 0.302 & 24.546 & 0.059 \\
853244 & 215.6125 & -0.394 & 27.121 & 0.029 & 0.436 & 27.355 & 0.034 & 0.498 & 26.295 & 0.054 & 0.319 & - & - \\
858989 & 215.6027 & -0.3992 & 25.717 & 0.035 & 0.196 & 26.016 & 0.036 & 0.239 & 24.769 & 0.034 & 0.164 & 23.52 & 0.035 \\
859454 & 215.5889 & -0.4063 & 26.043 & 0.014 & 0.304 & 26.292 & 0.019 & 0.351 & 25.184 & 0.025 & 0.231 & - & - \\
859464 & 215.5917 & -0.4049 & 26.866 & 0.019 & 0.306 & 27.102 & 0.025 & 0.353 & 26.031 & 0.036 & 0.228 & - & - \\
871337 & 215.6125 & -0.3949 & 25.846 & 0.027 & 0.208 & 25.971 & 0.038 & 0.229 & 25.191 & 0.04 & 0.135 & - & - \\
874062 & 215.6038 & -0.3994 & 27.225 & 0.03 & 0.269 & 27.424 & 0.043 & 0.306 & 26.45 & 0.068 & 0.198 & - & - \\
886741 & 215.6152 & -0.3943 & 27.512 & 0.034 & 0.173 & 27.805 & 0.035 & 0.212 & 26.571 & 0.034 & 0.143 & - & - \\
889136 & 215.5984 & -0.4029 & 26.991 & 0.022 & 0.244 & 27.288 & 0.022 & 0.293 & 26.049 & 0.023 & 0.2 & 24.795 & 0.032 \\
892554 & 215.6117 & -0.3963 & 27.154 & 0.047 & 0.3 & 27.46 & 0.043 & 0.353 & 26.2 & 0.056 & 0.237 & 24.93 & 0.075 \\
905168 & 215.6032 & -0.4013 & 26.533 & 0.035 & 0.262 & 26.751 & 0.047 & 0.301 & 25.727 & 0.04 & 0.195 & - & - \\
912240 & 215.6097 & -0.3984 & 27.281 & 0.065 & 0.35 & 27.379 & 0.072 & 0.385 & 26.681 & 0.114 & 0.23 & 25.999 & 0.244 \\
918325 & 215.6105 & -0.3982 & 26.659 & 0.018 & 0.378 & 26.885 & 0.02 & 0.432 & 25.842 & 0.033 & 0.277 & 24.796 & 0.072 \\
927325 & 215.6134 & -0.3973 & 27.113 & 0.028 & 0.291 & 27.418 & 0.028 & 0.345 & 26.163 & 0.029 & 0.233 & 24.886 & 0.037 \\
938088 & 215.6156 & -0.3967 & 27.187 & 0.034 & 0.246 & 27.487 & 0.038 & 0.295 & 26.24 & 0.032 & 0.2 & - & - \\
976489 & 215.6084 & -0.4023 & 27.078 & 0.024 & 0.208 & 27.356 & 0.028 & 0.25 & 26.165 & 0.038 & 0.169 & - & - \\
979358 & 215.6001 & -0.4067 & 27.343 & 0.033 & 0.179 & 27.615 & 0.04 & 0.217 & 26.437 & 0.062 & 0.144 & - & - \\
981628 & 215.6116 & -0.4009 & 26.472 & 0.016 & 0.263 & 26.732 & 0.017 & 0.309 & 25.595 & 0.034 & 0.203 & - & - \\
1003917 & 215.6038 & -0.406 & 26.428 & 0.018 & 0.287 & 26.733 & 0.018 & 0.339 & 25.476 & 0.02 & 0.228 & - & - \\
1015181 & 215.6113 & -0.4028 & 26.043 & 0.012 & 0.253 & 26.26 & 0.016 & 0.292 & 25.236 & 0.019 & 0.187 & 24.229 & 0.047 \\
1023938 & 215.6121 & -0.4028 & 26.433 & 0.02 & 0.289 & 26.604 & 0.029 & 0.327 & 25.706 & 0.041 & 0.204 & - & - \\
1031574 & 215.6126 & -0.4029 & 25.969 & 0.016 & 0.253 & 26.223 & 0.022 & 0.295 & 25.099 & 0.035 & 0.194 & 23.983 & 0.085 \\
1038160 & 215.6101 & -0.4045 & 26.795 & 0.038 & 0.259 & 26.928 & 0.046 & 0.287 & 26.129 & 0.063 & 0.174 & - & - \\
1045353 & 215.6116 & -0.4041 & 26.64 & 0.016 & 0.427 & 26.838 & 0.021 & 0.485 & 25.875 & 0.032 & 0.309 & - & - \\
1055460 & 215.6116 & -0.4045 & 27.006 & 0.045 & 0.296 & 27.244 & 0.059 & 0.342 & 26.167 & 0.063 & 0.222 & - & - \\
1073816 & 215.6089 & -0.4068 & 26.942 & 0.021 & 0.253 & 27.242 & 0.022 & 0.303 & 25.995 & 0.021 & 0.205 & 24.735 & 0.027 \\
\enddata


\tablecomments{The ID and the coordinates are those of \citetalias{Hoffmann2016}. LP: F350LP, V: F555W, I: F814W, H:F160W. Note that not all the Cepheids that are identified in the optical bands are detected in the H band.}


\end{deluxetable*}

\section{Data Files Obtained From MAST}\label{appendix_datafiles}
We list the information about all the data files we retrieved from the MAST database in Table \ref{tab:all_bands_fits_files}. 

\begin{deluxetable*}{cccccccccccc}


\tabletypesize{\scriptsize}


\tablecaption{The Observation ID of the F555W, F350LP, F814W, and F160W bands data files obtained from the MAST database, their exposure times in second (EXP), and their observation date. \label{tab:all_bands_fits_files}}


\tablehead{\colhead{ID} & \colhead{FILTER} & \colhead{EXP} & \colhead{DATE} & \colhead{ID} & \colhead{FILTER} & \colhead{EXP} & \colhead{DATE} & \colhead{ID} & \colhead{FILTER} & \colhead{EXP} & \colhead{DATE}} 

\startdata
ib1f25zkq & F555W & 600 & 2010-01-08 & ib1f36agq & F555W & 600 & 2010-03-08 & ib1f36amq & F350LP & 625 & 2010-03-08 \\
ib1f25zlq & F555W & 600 & 2010-01-08 & ib1f36aiq & F555W & 600 & 2010-03-08 & ib1f36aoq & F350LP & 625 & 2010-03-08 \\
ib1f25znq & F555W & 600 & 2010-01-08 & ib1f36akq & F555W & 600 & 2010-03-08 & ib1f39ipq & F350LP & 625 & 2010-03-14 \\
ib1f25zpq & F555W & 600 & 2010-01-08 & ib1f39ieq & F555W & 600 & 2010-03-14 & ib1f39irq & F350LP & 625 & 2010-03-14 \\
ib1f25zrq & F555W & 600 & 2010-01-08 & ib1f39ifq & F555W & 600 & 2010-03-14 & ib1f40zkq & F350LP & 625 & 2010-03-19 \\
ib1f25ztq & F555W & 600 & 2010-01-08 & ib1f39ihq & F555W & 600 & 2010-03-14 & ib1f40zmq & F350LP & 625 & 2010-03-19 \\
ib1f38coq & F555W & 600 & 2010-01-30 & ib1f39ijq & F555W & 600 & 2010-03-14 & ib1f0cvrq & F350LP & 625 & 2010-04-09 \\
ib1f38cpq & F555W & 600 & 2010-01-30 & ib1f39ilq & F555W & 600 & 2010-03-14 & ib1f0cvtq & F350LP & 625 & 2010-04-09 \\
ib1f38crq & F555W & 600 & 2010-01-30 & ib1f39inq & F555W & 600 & 2010-03-14 & ib1f0detq & F350LP & 625 & 2010-04-19 \\
ib1f38ctq & F555W & 600 & 2010-01-30 & ib1f40z9q & F555W & 600 & 2010-03-19 & ib1f0devq & F350LP & 625 & 2010-04-19 \\
ib1f38cvq & F555W & 600 & 2010-01-30 & ib1f40zaq & F555W & 600 & 2010-03-19 & ib1f31f1q & F814W & 600 & 2010-02-06 \\
ib1f38cxq & F555W & 600 & 2010-01-30 & ib1f40zcq & F555W & 600 & 2010-03-19 & ib1f31f3q & F814W & 600 & 2010-02-06 \\
ib1f31dsq & F555W & 600 & 2010-02-05 & ib1f40zeq & F555W & 600 & 2010-03-19 & ib1f31f5q & F814W & 600 & 2010-02-06 \\
ib1f31dtq & F555W & 600 & 2010-02-05 & ib1f40zgq & F555W & 600 & 2010-03-19 & ib1f31f7q & F814W & 600 & 2010-02-06 \\
ib1f31dvq & F555W & 600 & 2010-02-05 & ib1f40ziq & F555W & 600 & 2010-03-19 & ib1f32v3q & F814W & 600 & 2010-02-11 \\
ib1f31dxq & F555W & 600 & 2010-02-05 & ib1f0ai3q & F555W & 600 & 2010-03-30 & ib1f32v5q & F814W & 600 & 2010-02-11 \\
ib1f31e2q & F555W & 600 & 2010-02-06 & ib1f0ai5q & F555W & 600 & 2010-03-30 & ib1f32v7q & F814W & 600 & 2010-02-11 \\
ib1f31e4q & F555W & 600 & 2010-02-06 & ib1f0ai7q & F555W & 600 & 2010-03-30 & ib1f32v9q & F814W & 600 & 2010-02-11 \\
ib1f32uoq & F555W & 600 & 2010-02-11 & ib1f0ai9q & F555W & 600 & 2010-03-30 & ib1f33qnq & F814W & 600 & 2010-02-17 \\
ib1f32upq & F555W & 600 & 2010-02-11 & ib1f0cvgq & F555W & 600 & 2010-04-09 & ib1f33qpq & F814W & 600 & 2010-02-17 \\
ib1f32urq & F555W & 600 & 2010-02-11 & ib1f0cvhq & F555W & 600 & 2010-04-09 & ib1f33qrq & F814W & 600 & 2010-02-17 \\
ib1f32utq & F555W & 600 & 2010-02-11 & ib1f0cvjq & F555W & 600 & 2010-04-09 & ib1f33qtq & F814W & 600 & 2010-02-17 \\
ib1f32uvq & F555W & 600 & 2010-02-11 & ib1f0cvlq & F555W & 600 & 2010-04-09 & ib1f34miq & F814W & 600 & 2010-02-24 \\
ib1f32uxq & F555W & 600 & 2010-02-11 & ib1f0cvnq & F555W & 600 & 2010-04-09 & ib1f34mkq & F814W & 600 & 2010-02-24 \\
ib1f33q8q & F555W & 600 & 2010-02-17 & ib1f0cvpq & F555W & 600 & 2010-04-09 & ib1f34mmq & F814W & 600 & 2010-02-24 \\
ib1f33q9q & F555W & 600 & 2010-02-17 & ib1f0degq & F555W & 590 & 2010-04-19 & ib1f34moq & F814W & 600 & 2010-02-24 \\
ib1f33qbq & F555W & 600 & 2010-02-17 & ib1f0deiq & F555W & 590 & 2010-04-19 & ib1f36aqq & F814W & 600 & 2010-03-08 \\
ib1f33qdq & F555W & 600 & 2010-02-17 & ib1f0delq & F555W & 590 & 2010-04-19 & ib1f36asq & F814W & 600 & 2010-03-08 \\
ib1f33qfq & F555W & 600 & 2010-02-17 & ib1f0denq & F555W & 590 & 2010-04-19 & ib1f36auq & F814W & 600 & 2010-03-08 \\
ib1f33qhq & F555W & 600 & 2010-02-17 & ib1f0depq & F555W & 590 & 2010-04-19 & ib1f36awq & F814W & 600 & 2010-03-08 \\
ib1f34m1q & F555W & 600 & 2010-02-24 & ib1f0derq & F555W & 590 & 2010-04-19 & ib1f40zoq & F814W & 600 & 2010-03-19 \\
ib1f34m2q & F555W & 600 & 2010-02-24 & ib1f25zvq & F350LP & 625 & 2010-01-08 & ib1f40zqq & F814W & 600 & 2010-03-19 \\
ib1f34m4q & F555W & 600 & 2010-02-24 & ib1f25zxq & F350LP & 625 & 2010-01-08 & ib1f40zsq & F814W & 600 & 2010-03-19 \\
ib1f34m6q & F555W & 600 & 2010-02-24 & ib1f38czq & F350LP & 625 & 2010-01-30 & ib1f40zuq & F814W & 600 & 2010-03-19 \\
ib1f34maq & F555W & 600 & 2010-02-24 & ib1f38d1q & F350LP & 625 & 2010-01-30 & ib1f41mlq  & F160W & 502.9 & 2010-04-04 \\
ib1f34mcq & F555W & 600 & 2010-02-24 & ib1f31e7q & F350LP & 625 & 2010-02-06 & ib1f41mmq  & F160W & 502.9 & 2010-04-04 \\
ib1f35i9q & F555W & 600 & 2010-03-01 & ib1f31e9q & F350LP & 625 & 2010-02-06 & ib1f41moq  & F160W & 502.9 & 2010-04-04 \\
ib1f35iaq & F555W & 600 & 2010-03-01 & ib1f32uzq & F350LP & 625 & 2010-02-11 & ib1f41mpq  & F160W & 502.9 & 2010-04-04 \\
ib1f35icq & F555W & 600 & 2010-03-01 & ib1f32v1q & F350LP & 625 & 2010-02-11 & ib1f41mrq  & F160W & 452.9 & 2010-04-04 \\
ib1f35ieq & F555W & 600 & 2010-03-01 & ib1f33qjq & F350LP & 625 & 2010-02-17 & ib1f42t0q  & F160W & 502.9 & 2010-04-15 \\
ib1f35igq & F555W & 600 & 2010-03-01 & ib1f33qlq & F350LP & 625 & 2010-02-17 & ib1f42t1q  & F160W & 502.9 & 2010-04-15 \\
ib1f35iiq & F555W & 600 & 2010-03-01 & ib1f34meq & F350LP & 625 & 2010-02-24 & ib1f42t3q  & F160W & 502.9 & 2010-04-15 \\
ib1f36abq & F555W & 600 & 2010-03-08 & ib1f34mgq & F350LP & 625 & 2010-02-24 & ib1f42t4q  & F160W & 502.9 & 2010-04-15 \\
ib1f36acq & F555W & 600 & 2010-03-08 & ib1f35ikq & F350LP & 625 & 2010-03-01 & ib1f42t6q  & F160W & 452.9 & 2010-04-15 \\
ib1f36aeq & F555W & 600 & 2010-03-08 & ib1f35imq & F350LP & 625 & 2010-03-01 &  &  &  &  \\
\enddata




\end{deluxetable*}




\end{document}